\documentclass[journal]{IEEEtran}

\ifCLASSINFOpdf
\else
\fi

\usepackage{cite}
\usepackage{amsmath,amssymb,amsfonts}
\usepackage{algorithmic}
\usepackage{graphicx}
\usepackage{textcomp}
\usepackage{url,hyperref,lineno,microtype}
\usepackage[onehalfspacing]{setspace}
\usepackage{amsthm,amsmath,amssymb}
\usepackage{array,makecell}
\usepackage{multirow}
\usepackage{algorithmic}
\usepackage{algorithm}
\usepackage{comment}
\usepackage[T1]{fontenc}
\usepackage{graphicx}

\begin{document}

\title{Blockchain in Environmental Sustainability Measures: a Survey}
\author{Maria-Victoria Vladucu, Hailun Wu, Jorge Medina, Khondaker M. Salehin, Ziqian Dong, Roberto Rojas-Cessa$^*$ \thanks{M-V. Vladucu, H. Wu and Z. Dong are with the Network and Innovation Laboratory, Department of Electrical and Computer Engineering, New York Institute of Technology, New York, NY, 10023. K.M. Salehin is with Department of Computer Science, University of Portland, Portland, OR
97203, USA, and J. Medina and R. Rojas-Cessa are with the Networking and Intelligent Systems Laboratory, Department of Electrical and Computer Engineering, New Jersey Institute of Technology, Newark, NJ 07102, email: rojas@njit.edu.}}



\maketitle

\begin{abstract}
Real and effective regulation of contributions to greenhouse gas emissions and pollutants requires unbiased and truthful monitoring. Blockchain has emerged not only as an approach that provides verifiable economical interactions but also as a mechanism to keep the measurement, monitoring, incentivation of environmental conservationist practices and enforcement of policy. Here, we present a survey of areas in what blockchain has been considered as a response to concerns on keeping an accurate recording of environmental practices to monitor levels of pollution and management of environmental practices. We classify the applications of blockchain into different segments of concerns, such as greenhouse gas emissions, solid waste, water, plastics, food waste, and circular economy, and show the objectives for the addressed concerns. We also classify the different blockchains and the explored and designed properties as identified for the proposed solutions. At the end, we provide a discussion about the niches and challenges that remain for future research.
\end{abstract}

\begin{IEEEkeywords}
Environmental sustainability, blockchain, pollution monitoring, plastic waste, greenhouse gas emissions, food waste, food security, waste water, environmental policies, circular economy.
\end{IEEEkeywords}

\IEEEpeerreviewmaketitle

\section{Introduction}
\IEEEPARstart{E}{nvironmental} sustainability is threatened by the increasing environmental pollution that stem from our daily activities and production practices \cite{kennish2002environmental}. Unsustainable use of natural resources and mismanagement of waste are all contributing factors that are manifesting in the predicted phenomena the scientific community has long been sounding the alarms on. Global warming, extreme weather events such as drought, flooding, loss of land due to sea level rise, threaten the production of food and health \cite{cook2018climate}. The industrial processes that support the consumption-driven economy are often the culprits that pollute the air, water, and soil damaging the environment and threatening lives of all species \cite{berry1998proactive}.

Many countries and regions around the world are joining forces to take action against climate change to cut down emissions and limit global warming to 1.5$^\circ$ by the end of the century through the Paris Agreement \cite{paris_agreement}. 
Halting the generation of polluters, while an ideal goal in the long term, might be currently infeasible as communities' economies depend on industrialization and diverse human activity. Therefore, achieving a balance in managing pollutants and their production emerges as a pragmatic solution \cite{usman2023mercosur}.

Critical tools for achieving this balance include effective approaches to monitor, manage, and control pollutants and resource utilization \cite{shrivastava1995role}. However, challenges persist in accurately measuring, tracking, and reporting waste generation and resource usage, particularly due to conflicting interests among stakeholders within the management chain. In such conditions, the reliability of data regarding resource status or waste generation, crucial for effective management, may be compromised \cite{schaffer2023conflicts}.

To tackle these challenges, an electronic recording system with features such as data availability, transparency, and protection against manipulation is essential. Recently, blockchain technology has emerged as a promising solution. Blockchain keeps data immutable by resorting to a distributed and parallel data ledger that makes it not only fault tolerant but also resistant to Byzantine attacks that threaten data integrity and trust \cite{allena2020blockchain}.
Blockchain offers many properties beyond its immutability property. Its distributed data-keeping system, the use of consensus bring trustworthiness and reliability that users and policymakers seek for securing data, processes, and transactions that record activities associated with environmental variables.

In this survey, we review existing research that applies blockchain technology to enhance trust in the recording, sensing, and management of agents affecting environmental sustainability. These agents include greenhouse gas (GHG) emissions, solid waste, plastic, food, water, and a new circular economy concept with the goal of reducing, reusing, and re-purposing waste to reduce their environmental impact. These works represent interdisciplinary approaches that leverage environmental, social, financial, and engineering solutions. We have excluded applications of blockchain technology in energy and chemical waste management \cite{mishra2019heavy} due to the broad scope of the former and the localized nature of the latter. 
Our objectives in this paper are to address the following questions: 1) What are the motivations for using blockchain technology in waste, food, water and circular-economy management? 2) What specific features of blockchain are sought in existing research? 3) What gaps exist in current research that warrant further investigation? 

Through exhaustive literature research, we aim to provide insights into these questions, highlight overlaps, and identify existing challenges across the surveyed work. Ultimately, our analysis seeks to guide future research directions in this evolving field.

We categorize literature work in these topics. Furthermore, we identify the blockchain frameworks used on the various works in each topic, their consensus approach, and the technologies used to support the collection of data, and remaining challenges for each categorized work.

For completeness, we also identify the existing surveys on the application of blockchain on the topics covered here, and discuss and compare their focus and features. We also indicate how our survey differentiates from those. At the end, we identify published work that provide source code for the implementation of blockchain artifacts, and existing challenges that point to future research.

The remainder of this survey is organized as follows. Section \ref{sec:features} presents an overview of blockchain features that make blockchain a technology of interest to support the monitoring of environmental and polluting variables. 
Section \ref{sec:blockchain-management} presents existing applications of blockchain on management of carbon, greenhouse gas emissions, solid waste, plastic, food waste, water use, and circular economy. 
Section \ref{sec:discussion} presents future challenges, and Section \ref{sec:conclusions} presents our conclusions.

\begin{figure}[htp!]
    \centering
\includegraphics[width=0.95\columnwidth]{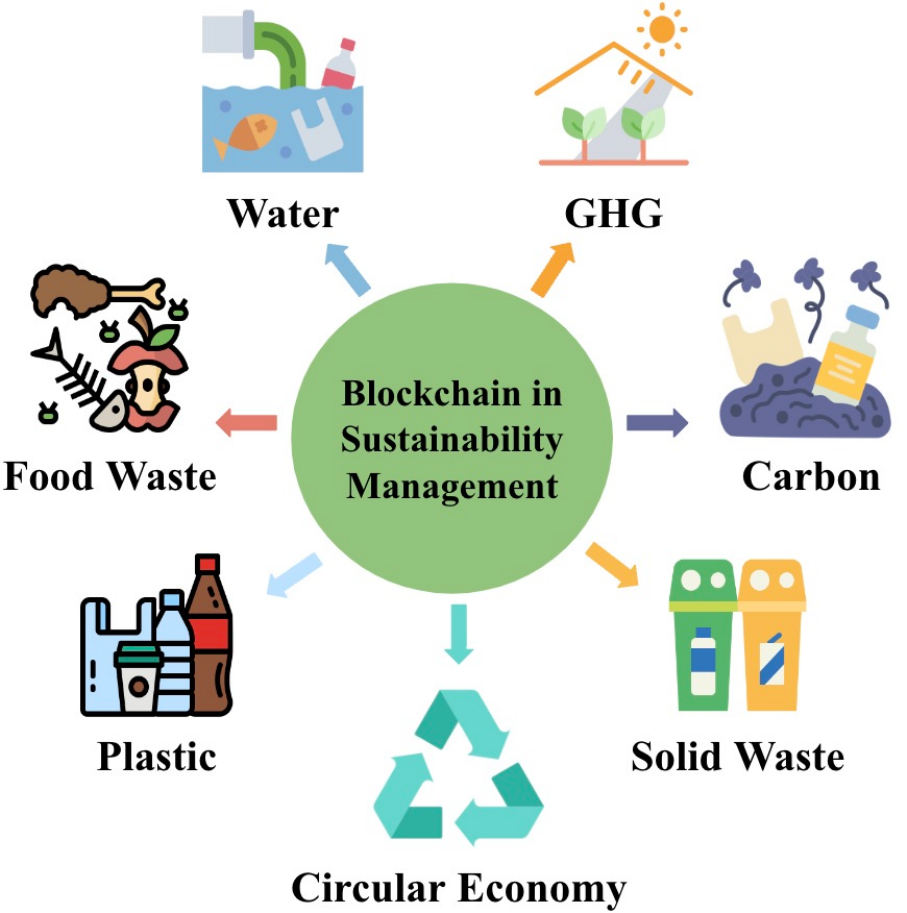}
    \caption{Content and organization of this paper: Applications of blockchain on the management of carbon, GHG emissions, solid waste, plastic waste, water, food waste, and circular economy.}
    ~\label{fig:overview}
\end{figure}

Figure \ref{fig:overview} shows the organization of this survey. 

\section{Blockchain}
\label{sec:features}

\begin{table*}[thb!]
  \centering
  \caption{Terminology used in this paper.}
   \renewcommand{\arraystretch}{1}
    \begin{tabular}{|@{}>{\centering\arraybackslash}p{2.5cm}@{}|@{}>{\centering\arraybackslash}m{5.2cm}@{}|@{}>{\centering\arraybackslash}p{2.5cm}@{}|@{}>{\centering\arraybackslash}p{0.8cm}@{}|@{}>{\centering\arraybackslash}p{0.8cm}@{}|@{}>{\centering\arraybackslash}p{0.8cm}@{}|@{}>{\centering\arraybackslash}p{0.8cm}@{}|@{}>{\centering\arraybackslash}p{0.8cm}@{}|@{}>{\centering\arraybackslash}p{0.8cm}@{}|@{}>{\centering\arraybackslash}p{0.8cm}@{}|}\hline

    \multicolumn{1}{|c|}{\multirow{2}{*}{\textbf{Category}}}&
    \multicolumn{1}{c|}{\multirow{2}{*}{\textbf{Term}}}&
    \multicolumn{1}{c|}{\multirow{2}{*}{\textbf{Abbreviation}}}&
    \multicolumn{7}{c|}{\textbf{Paper Section}}\\\cline{4-10}

    & & &
    \textbf{\scriptsize GHG}&
    \textbf{\scriptsize C}&
    \textbf{\scriptsize SW}&
    \textbf{\scriptsize PW}&
    \textbf{\scriptsize FW}&
    \textbf{\scriptsize WM}&
    \textbf{\scriptsize CE}
    \\
    \hline

    \multirow{10}{*}{\makecell{\\Consensus \\algorithms}} & 
    {Proof of Work} & PoW &
    {\scriptsize \checkmark}&
    {}&
    {\scriptsize \checkmark}&
    {\scriptsize \checkmark}&  
    {}&
    {\scriptsize \checkmark}&
    {} \\\cline{2-10}

   & {Proof of Useful Work} & PoUW &
    {}&
    {\scriptsize \checkmark}&
    {}&
    {}&  
    {}&
    {}&
    {} \\\cline{2-10}
    &{Delegated Proof of Stake} & DPoS &
    {}&
    {}&
    {}&
    {}&  
    {}&
    {\scriptsize \checkmark}&
    {} \\\cline{2-10}
     &{Practical Byzantine Fault Tolerance } & PBFT &
    {}&
    {\scriptsize \checkmark}&
    {}&
    {\scriptsize \checkmark}&  
    {\scriptsize \checkmark}&
    {}&
    {} \\\cline{2-10}
     &{Raft} & \textemdash   &
    {}&
    {\scriptsize \checkmark}&
    {}&
    {}&  
    {}&
    {}&
    {} \\\cline{2-10}
    &{Kafka}& \textemdash  &
    {}&
    {\scriptsize \checkmark}&
    {}&
    {}&  
    {}&
    {\scriptsize \checkmark}&
    {} \\\cline{2-10}
     &{YetAnotherConsensus} & YAC &
    { }&
    {\scriptsize \checkmark}&
    {}&
    {}&  
    {}&
    {}&
    {} \\ \cline{2-10}

     &{Proof of Randomness} & PoR &
    {}&
    {\scriptsize \checkmark}&
    {}&
    {}&  
    {}&
    {}&
    {} \\\cline{2-10}
    
     &{Delegated Proof of Reputation} & DPoR &
    {}&
    {\scriptsize \checkmark}&
    {}&
    {}&  
    {}&
    {}&
    {} \\\cline{2-10}

    &{Proof of Authority } & PoA &
    {}&
    { }&
    {}&
    {}&  
    {}&
    {\scriptsize \checkmark}&
    {} \\\cline{2-10}
     &{Proof of Authentication } & PoAh &
    {}&
    {}&
    {}&
    {}&  
    {}&
    {\scriptsize \checkmark}&
    {} \\\cline{2-10}
    &{Proof-of-single-learning  } & PoSL &
    {}&
    {}&
    {}&
    {}&  
    {}&
    {\scriptsize \checkmark}&
    {} \\\cline{2-10}
    &{Proof-of-multiple-learning} &  PoML &
    {}&
    {}&
    {}&
    {}&  
    {}&
    {\scriptsize \checkmark}&
    {} \\\cline{2-10}
     &{Proof of Stake Time } & PoST &
    { }&
    {}&
    {}&
    {}&  
    {}&
    {\scriptsize \checkmark}&
    {} \\\cline{2-10}
      &{Proof-of-trust } & PoT &
    { }&
    {}&
    {}&
    {}&  
    {}&
    {\scriptsize \checkmark}&
    {} \\\cline{2-10}
      &{Proof of Vote } & PoV &
    { }&
    {}&
    {}&
    {}&  
    {}&
    {\scriptsize \checkmark}&
    {} 
  \\\hline

    \multirow{6}{*}{\makecell{Frameworks}}& 
    {Bitcoin} & \textemdash  &
    {}&
    {\scriptsize \checkmark}&
    {}&
    {}&  
    {}&
    {}&
    {} \\\cline{2-10}
   &  {Ethereum} & \textemdash  &
    {\scriptsize \checkmark}&
    {\scriptsize \checkmark}&
    {\scriptsize \checkmark}&
    {\scriptsize \checkmark}&  
    {}&
    {\scriptsize \checkmark}&
    {} \\\cline{2-10}
    &{Hyperledger Fabric } & \textemdash  &
    {\scriptsize \checkmark }&
    {\scriptsize \checkmark}&
    {\scriptsize \checkmark}&
    {\scriptsize \checkmark}&  
    {\scriptsize \checkmark}&
    {\scriptsize \checkmark}&
    {\scriptsize \checkmark} \\\cline{2-10}
    &{Hyperledger Sawtooth  } & \textemdash  &
    {}&
    {}&
    {}&
    {}&  
    {\scriptsize \checkmark}&
    {}&
    {} \\\cline{2-10}
    &{Hyperledger Iroha } &\textemdash   &
    {}&
    {\scriptsize \checkmark}&
    {}&
    {}&  
    {}&
    {}&
    {}  \\\cline{2-10}
    &{Corda} & \textemdash   &
    {}&
    {}&
    {}&
    {}&  
    {}&
    {\scriptsize \checkmark}&
    {} 
  \\\hline
    
 \multirow{10}{*}{\makecell{Features }}& 
    {Anonymity} & AN &
    {\scriptsize \checkmark}&
    {\scriptsize \checkmark}&
    {\scriptsize \checkmark}&
    {\scriptsize \checkmark}&  
    {\scriptsize \checkmark}&
    {\scriptsize \checkmark}&
    {\scriptsize \checkmark} \\\cline{2-10}
   & {Immutability } & IM &
    {\scriptsize \checkmark}&
    {\scriptsize \checkmark}&
    {\scriptsize \checkmark}&
    {\scriptsize \checkmark}&  
    {\scriptsize \checkmark}&
    {\scriptsize \checkmark}&
    {\scriptsize \checkmark} \\\cline{2-10}
    &{Smart contracts} & SC &
    {\scriptsize \checkmark}&
    {\scriptsize \checkmark}&
    {\scriptsize \checkmark}&
    {\scriptsize \checkmark}&  
    {\scriptsize \checkmark}&
    {\scriptsize \checkmark}&
    {\scriptsize \checkmark} \\\cline{2-10}
    &{Data Governance} & DG &
    {}&
    {}&
    {\scriptsize \checkmark}&
    {}&  
    {}&
    {}&
    {\scriptsize \checkmark} \\\cline{2-10}
    &{Data Security} &  DS &
    {}&
    {\scriptsize \checkmark}&
    {\scriptsize \checkmark}&
    {}&  
    {\scriptsize \checkmark}&
    {\scriptsize \checkmark}&
    {} \\\cline{2-10}
    &{Incentives} & IN &
    {\scriptsize \checkmark}&
    {\scriptsize \checkmark}&
    {\scriptsize \checkmark}&
    {\scriptsize \checkmark}&  
    {}&
    {}&
    {} \\\cline{2-10}
     &{ Payments} & PY &
    {}&
    {\scriptsize \checkmark}&
    {}&
    {}&  
    {}&
    {\scriptsize \checkmark}&
    {} \\\cline{2-10}
     &{Traceability} & TC   &
    {\scriptsize \checkmark}&
    {\scriptsize \checkmark}&
    {\scriptsize \checkmark}&
    {\scriptsize \checkmark}&  
    {\scriptsize \checkmark}&
    {\scriptsize \checkmark}&
    {\scriptsize \checkmark} \\\cline{2-10}
    &{Tokenization}& TK &
    {}&
    {\scriptsize \checkmark}&
    {}&
    {}&  
    {}&
    {\scriptsize \checkmark}&
    {\scriptsize \checkmark} \\\cline{2-10}
     &{Transparency} & TP &
    {\scriptsize \checkmark}&
    {\scriptsize \checkmark}&
    {\scriptsize \checkmark}&
    {\scriptsize \checkmark}&  
    {\scriptsize \checkmark}&
    {\scriptsize \checkmark}&
    {\scriptsize \checkmark} \\\cline{2-10}
    &{Tracking}& TR &
    {}&
    {\scriptsize \checkmark}&
    {\scriptsize \checkmark}&
    {\scriptsize \checkmark}&  
    {\scriptsize \checkmark}&
    {\scriptsize \checkmark}&
    {}
   \\\hline
    
    \multirow{15}{*}{\makecell{\\Additional  \\technology }}& 
    {Artificial Intelligence} & AI &
    {}&
    {\scriptsize \checkmark}&
    {}&
    {\scriptsize \checkmark}& 
    {}&
    {}&
    {} \\\cline{2-10}
    &{Vehicular ad hoc network  } & VANET &
    {}&
    {\scriptsize \checkmark}&
    {\scriptsize \checkmark}&
    {}&  
    {}&
    {}&
    {} \\\cline{2-10}
   
    &{InterPlanetary File System } & IPFS &
    {\scriptsize \checkmark}&
    {}&
    {\scriptsize \checkmark}&
    {}&  
    {}&
    {\scriptsize \checkmark}&
    {} \\\cline{2-10}
    
    &{Internet of Things} & IoT &
    {}&
    {}&
    {\scriptsize \checkmark}&
    {\scriptsize \checkmark}&  
    {\scriptsize \checkmark}&
    {\scriptsize \checkmark}&
    {\scriptsize \checkmark} \\\cline{2-10}
  
      &{ Internet of Agricultural Things} & IoAT &
    { }&
    {}&
    {}&
    {}&  
    {}&
    {\scriptsize \checkmark}&
    {} \\\cline{2-10}
      &{Internet of Underwater Things } &IoUT  &
    { }&
    {}&
    {}&
    {}&  
    {}&
    {\scriptsize \checkmark}&
    {} \\\cline{2-10}
    &{ Industrial Internet of Things} & IIoT &
    { }&
    {}&
    {}&
    {}&  
    {}&
    {\scriptsize \checkmark}&
    {} \\\cline{2-10}
      &{Industrial Internet of Water Things} &IIoWT &
    { }&
    {}&
    {}&
    {}&  
    {}&
    {\scriptsize \checkmark}&
    {} \\\cline{2-10}
   &  {Quick-response Code} & QR Code&
    {}&
    {}&
    {\scriptsize \checkmark}&
    {}&  
    {\scriptsize \checkmark}&
    {}&
    {} \\\cline{2-10}
    
     
      &{Ultra-high-temperature processing } & UHT &
    { }&
    {}&
    {\scriptsize \checkmark}&
    {}&  
    {}&
    {}&
    {} \\\cline{2-10}
      &{Radio Frequency Identification } & RFID &
    { }&
    {}&
    {\scriptsize \checkmark}&
    {}&  
    {\scriptsize \checkmark}&
    {\scriptsize \checkmark}&
    {\scriptsize \checkmark} \\\cline{2-10}
   
      &{ Decentralized Application} & DApp &
    { }&
    {}&
    {}&
    {\scriptsize \checkmark}&  
    {}&
    {}&
    {} \\\cline{2-10}
      &{Machine Learning } & ML &
    { }&
    {}&
    {}&
    {}&  
    {\scriptsize \checkmark}&
    {\scriptsize \checkmark}&
    {\scriptsize \checkmark} \\\cline{2-10}
      &{User Interface } & UI &
    { }&
    {}&
    {}&
    {}&  
    {}&
    {\scriptsize \checkmark}&
    {} \\\cline{2-10}
      &{Wireless Sensor Network } & WSN  &
    { }&
    {}&
    {}&
    {}&  
    {}&
    {\scriptsize \checkmark}&
    {} 
  \\\hline
   



   
  
    \end{tabular}%
  \label{tab:term2}%
\end{table*}%

Blockchain is a decentralized and distributed digital ledger that keeps data immutable to safeguard data and the record-keeping process. Blockchain operates across a peer-to-peer (P2P) network of nodes, called miners or validators, that interplay a consensus algorithm to determine data as truthful. Data is recorded after consensus is affirmatively verified. The distributed ledger is organized as blocks of verified transactions. These blocks are linked as a chain to provide historical immutable records. With the combination of distributed operations of the consensus algorithm and cryptographic schemes, blockchain makes it difficult for adversaries to tamper with the information stored in the distributed ledger.

Data stored in the blockchain represent verified client transactions. Authenticated clients issue transactions representing operations on objects or with other clients. These transactions could be the transfer of ownership of a physical or digital object that becomes the record of new data. Such data is recorded, updating the state of the blockchain. Validator nodes run a consensus algorithm that executes rules. These rules are required to reach agreements for adding new blocks to keep the distributed ledger consistent. Data can also be recorded as smart contracts, which are pieces of code that are executed on the blockchain for the correct execution of a transaction. A smart contract is an artifact that determines whether the operations of a transaction are completed. When combined with the immutability property of a blockchain, it becomes a trusted digital arbitration mechanism.

\subsection{Blockchain framework and consensus}
A {\it public blockchain} is a decentralized blockchain with tamper-proof features that allow any user to join the network, read the blockchain's content, submit new transactions or verify the correctness of the blocks, and participate in the consensus processes. Well-known examples of public blockchains are Bitcoin \cite{bitcoin}, NXT \cite{king2012ppcoin}, and Ethereum \cite{ethereum}. 

A {\it private blockchain} uses an entity as the sole party for granting users permission to join the network and write or send transactions to the blockchain. Well-known examples of private blockchains are Hyperledger Fabric, Ripple, and Eris \cite{e-voting}.

A {\it consortium blockchain} is a permissioned blockchain in which participants must be granted access by the consortium. It follows a hierarchical model managed by a group of entities that join the consortium rather than a sole entity~\cite{consortium}.

A {\it blockchain framework} is a set of tools that enables the development of blockchain-based applications. The frameworks support the function of consensus mechanisms, smart contracts, cryptographic functions, and APIs to build decentralized applications (dApps) or blockchain-based solutions.

{\it Bitcoin} is an open-source P2P network that uses blockchain for payment exchange and cyber currency generation \cite{bitcoin}. This public blockchain handles between 4.6 and 7 transactions per second, making it hardly scalable. 

{\it Ethereum} uses a P2P network with Proof of Work (PoW) to handle up to 15 transactions per second, thus having low scalability. Unlike Bitcoin, Ethereum is a programmable blockchain where users can build and deploy decentralized applications on its P2P network \cite{ethereum}.


{\it Hyperledger} is an open-source blockchain platform hosted by the Linux Foundation’s Hyperledger project designed to meet confidentiality, privacy, and scalability requirements while maintaining a global collaboration between finance, banking, supply chains, manufacturing, and technology \cite{hyperledger_foundation}. Hyperledger is represented by three main frameworks: Hyperledger Fabric~\cite{hyperledger_foundation}, Sawtooth~\cite{HS}, and Iroha~\cite{HIroha}.
Hyperledger Fabric is a consortium blockchain that uses consensus algorithms such as  Practical Byzantine Fault Tolerance (PBFT), Raft, and Kafka. It uses a modular architecture where only designated users can access the data \cite{hyperledger_foundation}.

Hyperledger Sawtooth~\cite{HS} is a framework for building enterprise-grade distributed ledgers that focus on security, scalability, and modularity. 
Hyperledger Sawtooth is reported to have reached end-of-life status on February 1, 2024~\cite{HS}.

Hyperledger Iroha~\cite{HIroha} is a distributed ledger for C++ developers. It provides a framework with a pre-defined set of commands, permissions, and queries that can be used with various client libraries (Java, Python, JavaScript, Swift) to easily create applications for desktop and mobile platforms. 

{\it Corda }is an open-source blockchain platform developed by R3. Corda is designed for enterprise environments, highlighting data privacy, security, and compliance. Corda shares data on a need-to-know basis because parties may be competitors who want to keep business relationships and details secret from one another. Participants must first obtain a digital certificate before joining the network~\cite{WT47, corda}.

A {\it consensus algorithm} is a protocol used by the nodes of a blockchain network to ensure consistency among the distributed ledgers~\cite{e-voting}.

{PoW} is a consensus algorithm used in Bitcoin and Ethereum \cite{e-voting} where nodes, called miners or validators, compete to solve a computationally challenging puzzle to decide who leads the linking of the new block to the last block in the valid blockchain. The winner is the miner who finds the correct solution first and gets the right to create a new block in the blockchain. The process is called mining~\cite{e-voting}. 

{\it Proof of Useful Work (PoUW)} is a variation of PoW. PoUW introduces the requirement that the work being done is not just computational but also useful in some real-world applications~\cite{pouw-c}. PoUW incentivizes miners to perform computational tasks that are beneficial outside of the blockchain, such as scientific research or data processing, in addition to securing the blockchain network~\cite{pouw}. 

{\it Proof of Stake (PoS)} is a consensus algorithm that eliminates the computational-intensive mining process used in PoW \cite{king2012ppcoin}. The miners are called forgers, and the process is known as forging. Forgers deposit a number of coins that they own as stakes. The protocol selects the next forger in the network according to the forger's stake. PoS uses forger selection methods: the coin-age selection and the randomized block selection \cite{ismail2019review}. 

{\it Delegated Proof of Stake (DPoS)} is a consensus algorithm similar to PoS which is proposed to resolve the ``the rich getting richer in PoS'' problem. In DPoS, the nodes in the network select delegates through voting to validate the blocks \cite{awalu2019development}. The process in DPoS is divided into two stages: 1) election witnessing, where transactions are verified, and 2) block generation. The more blockchain stakes they have, the higher their possibility of being a forger. The identity of the forgers is already known and static. However, this process is vulnerable to collusion attacks \cite{yang2019delegated}.

{\it Practical Byzantine Fault Tolerance (PBFT)} is a consensus algorithm proposed to solve the Byzantine Generals Problem and optimizes for low overhead time to solve problems associated with already available Byzantine Fault Tolerance solutions. This game theory problem is designed to work efficiently in asynchronous systems~\cite{awalu2019development}. 

{\it Raft} and {\it Kafka} are two crash fault-tolerant consensus mechanisms used with Hyperledger. Raft is a consensus algorithm for managing a leader-follower dynamic in which the leader is elected and replicates messages, called logs, to the followers. It separates key elements such as leader election, log replication, and safety~\cite{raft}. Kafka is a crash fault-tolerant consensus and also uses the ``leader and follower” configuration as the one used in Raft. Transactions are replicated from the leader node to the follower nodes. If the leader node goes down, one of the followers becomes the leader, and ordering can continue, ensuring fault tolerance~\cite{kafka}.

{\it YetAnotherConsensus} is a consensus algorithm used with Hyperledger Iroha \cite{HIroha} to address the challenges of ineffective message transmission and dominance of leaders commonly encountered in traditional Byzantine fault-tolerant consensus
algorithm~\cite{yac}. 

{\it Proof of Randomness (PoR)} is a low-energy-cost consensus mechanism in which each node uses a true random number generator and hash algorithm to fulfill the PoR protocol. Unlike PoW or PoS, PoR can generate a block without competition in computing power or cryptocurrency stakes~\cite{por,kim2020blockchain_CM1}. 

{\it Delegated Proof of Reputation (DPoR}) is a consensus mechanism where miners are selected based on their reputation. It has four stages: voting power calculation, voting and calculation of the Reputation of Miner (RoM), block packing and propagation, and block verification and appending. Overall, DPoR's security is improved because DPoS performs better than the enhanced DPoS at finding malicious miners faster~\cite{hu2020delegated_CM17}.

{\it Proof of Authority (PoA)} is a permissioned consensus algorithm similar to BFT and is used to reduce resource consumption where authorized signers can create new blocks at their discretion~\cite{WT3}. The validators are restricted to a fixed set of $n$ authorized nodes called sealers. The reputation of the sealer is at stake. Malicious behaviors damage their reputation. Sealers should be incentivized to retain their reputation or gain it subsequently, as this will prevent sealers from associating with any malicious activity on the network. PoA depends on a mining rotation mechanism, so other sealers in the network strictly monitor the actions of different sealers~\cite{poah}.

{\it Proof of Authentication (PoAh)} is a consensus algorithm that uses distributed authentication to reduce the energy consumption of those that use centralized methods~\cite{WT27}.

{\it Proof-of-Single-Learning (PoSL)} is a consensus mechanism where one blockchain node is selected at
random to validate the data with a machine learning (ML) model, while {\it Proof-of-Multiple-Learning (PoML)} uses several validators~\cite{WT57}.

{\it Proof of Stake Time (PoST)} is a consensus mechanism that solves the problem of ``the coin-age." Validators are selected based on their stake in the network. They are required to demonstrate consistent participation and contribution over time to be chosen to create new blocks.

{\it Proof-of-trust (PoT)} is a consensus algorithm that selects validators based on the participants’ trust values. PoT has a centralized reputation-based approach to reach a consensus and avoids low throughput and high resource consumption. It has a better transaction rate than PoS or PoW.
By separating participants' powers in the consensus process, PoT promotes fairness and security~\cite{pot}.

{\it Proof of Vote (PoV)} is a consensus protocol based on a voting mechanism and consortium blockchain. PoV separates voting rights and executive rights. The consensus mimics the voting campaign by designing four types of network participants: commissioner, butler, butler candidates, and ordinary participants. 
Ordinary participants can join and exit the network to vote without administrative rights~\cite{li2017proof}.


\subsection{Blockchain Features in Environmental Monitoring}

Blockchain has the potential to revolutionize the management of environmental variables, especially those that damage the environment, by securely recording and providing access to information. Blockchain provides features such as data security, transparency, and trust, enhancing efficiency across diverse industries and applications. Some blockchain features are inherent attributes, like data immutability, and others are features enabled by the use of smart contracts in a blockchain. For example, a smart contract can be deployed on a blockchain to facilitate the establishment of environmental policy and ensure enforcement. Figure~\ref{blockchainFeatures} shows blockchain features mostly sought after in environmental management and monitoring applications. These features are described as follows:
\begin{figure}[htp!]
    \centering
\includegraphics[width=\columnwidth]{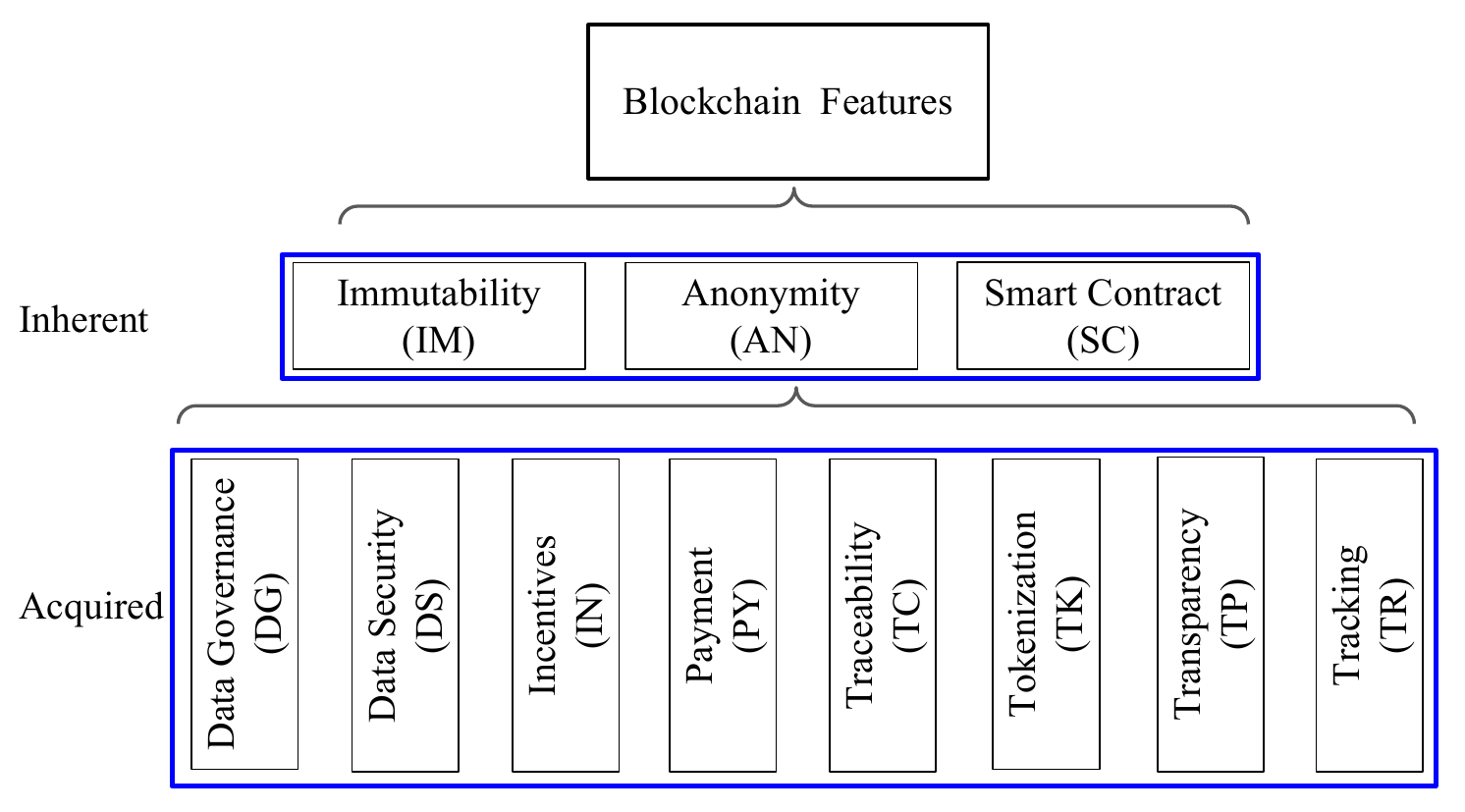}
    \caption{Blockchain features in environmental monitoring: inherent ones, and acquired ones, identified in this paper.}
    ~\label{blockchainFeatures}
\end{figure}

\noindent {\it Immutability~(IM)} is the characteristic that implies that transactions stored in the blockchain cannot be tampered with. Once data is recorded in the blockchain, the data becomes tamper-proof. This measure guarantees the integrity of the data within the blockchain, as any attempts to modify it are infeasible.

\noindent {\it Anonymity~(AN)} is the attribute of a system where participants' and stakeholders' information is protected and kept private.

\noindent {\it Smart contracts (SC)}:
A smart contract~\cite{buterin2014next} is a self-executing trusted code that runs on a blockchain network without needing a trusted or centralized node. Blockchain clients issue transactions to trigger smart contracts to perform functions on the blockchain. 

Other acquired features enabled by blockchain include the following:

\begin{itemize}

\item {\it Data Governance~(DG)} defines how data is shared and who or what processes have access to the recorded data. 
   
\item {\it Data Security~(DS)} is the property of a system that is immune to either a specific or a group of attacks against the user data.
   
\item {\it Incentives~(IN)} refers to mechanisms designed to encourage user participation. These incentives encompass both rewards and penalties within the system. Rewards are employed to promote favorable behavior and adherence to the established rules defined by smart contracts, while penalties serve to discourage unfavorable behaviors.   
   
    
    \item {\it Payments~(PY)} is the process of \textit{transferring} cryptocurrencies or tokens between blockchain clients' accounts in exchange for goods or services. 

    \item {\it Traceability~(TC)} is the ability to access chronological information of a client, a physical or digital object, or a process through recorded transactions. Traceability allows users to follow the history of an object. Provenance, as a specialized application of tracking, uses traceability to identify the origin of the subject.

     \item {\it Tokenization~(TK)} is the representation of a universal value of physical or digital assets, or ownership rights on a blockchain network.
    Tokens can be exchanged or generated in a blockchain as a result of a smart contract. 
    \item {\it Transparency~(TP)} is the property that allows users to access information recorded in the blockchain for verification. Users may look into this feature to access historical data.
    
     \item {\it Tracking~(TR)} is the feature of blockchain that allows access to the data regarding the current location or status of objects or processes. 
   

\end{itemize}

Table \ref{tab:term2} outlines the terminologies and abbreviations used in this survey. They are categorized based on the blockchain framework, features, consensus, additional technologies, and where they appear in the sections (greenhouse gas (GHG) emissions, carbon (C), solid waste (SW), plastic waste (PW), food waste (FW), water management (WM), and circular economy (CE)) of the survey for easier reference.

\section{Blockchain-based Management of Critical Elements for Sustainability}
\label{sec:blockchain-management}

In this section, we overview the existing blockchain-based approaches to improve the management of greenhouse gas emissions and carbon, solid waste, plastic waste, food waste, water use, and circular economy. We classify existing work by the objective and motivation of using blockchain in the management the variable that impacts environmental sustainability. We highlight the features that motivates the adoption of blockchain and the stakeholders that have been identified in the literature that may have a conflict of keeping data immutable. The classification also highlights the used blockchain framework and the supporting technologies. Each of these classifications ends by presenting the outstanding challenges in each application that still are open for research.

\subsection{Greenhouse Gas Emissions}
\label{sec:greenhouse-emissions}

GHG are those that contribute to the warming of the Earth's atmosphere. These gases include carbon dioxide (CO$_2$), methane (CH$_4$), nitrous oxide (N$_2$O), hydrofluorocarbons (HFCs), perfluorocarbons (PFCs), sulfur hexafluoride (SF$_6$), and nitrogen trifluoride (NF$_3$) \cite{EPSGHG, GW18}. GHG emissions also include various air pollutants, such as particulate matter PM${2.5}$, PM${5}$, PM${10}$, and volatile organic compounds \cite{EPSGHG}.
\begin{figure}[htp!]
    \centering
\includegraphics[width=0.9\columnwidth]{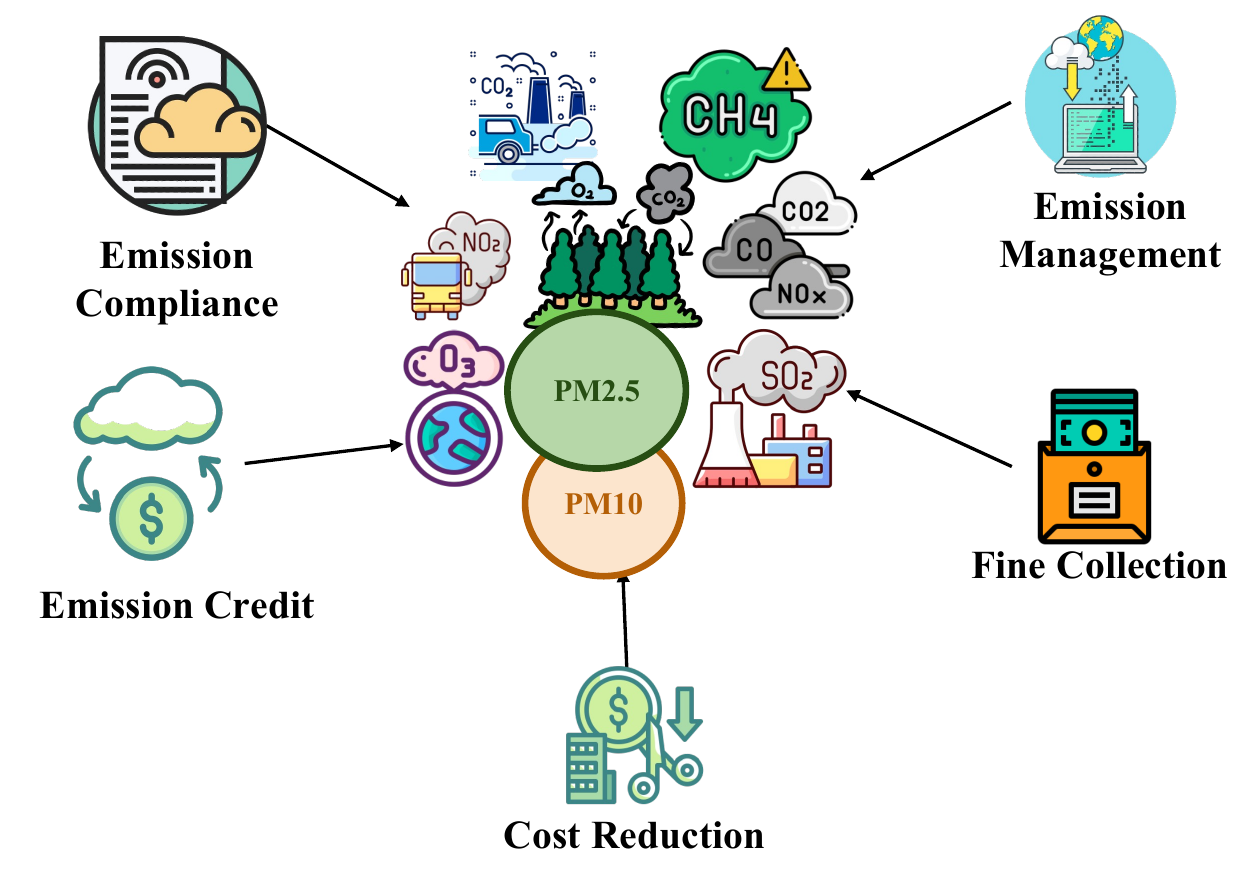}
    \caption{Objectives pursued in reported blockchain applications for management of GHG emissions.}
    ~\label{fig:ghg}
\end{figure}

Carbon dioxide is primarily generated by burning material and also by chemical reactions. As the most prevalent GHG, it is a major concern. Methane emissions stem from various sources, including coal, natural gas, and oil production, as well as livestock and agricultural practices, land use changes, and the decomposition of organic waste in municipal solid waste \cite{EPSGHG}. N$_2$O is produced by industrial activities, fossil fuel combustion, agricultural practices, water treatment, and solid waste management. Fluorinated gases, emitted by households, industrial processes, and products, are also significant contributors to GHG emissions. These gases are classified as having a high global warming potential (GWP) because they deplete ozone in the atmosphere and trap high concentrations of CO$_2$.

GHG emissions are released from any process involving combustion. Some emissions are generated directly or indirectly \cite{GW11}. Vehicles, including cars, are significant and direct contributors of these gases. Building of batteries may be an indirect contribution of GHGs. Monitoring the generation of GHG emissions is recognized as a critical need, especially given the widespread use of internal-combustion-engine vehicles. Therefore, monitoring efforts must encompass not only individuals but also vehicles. To ensure effective monitoring, it is essential to equip internal-combustion-engine vehicles with monitors, and the data collected must be protected from tampering. Industries and households that utilize combustion in various capacities are another significant source of GHG emissions that require monitoring.

In this section, we examine research on monitoring and recording GHG emissions data, identifying the objectives that drive the implementation of blockchain technology, the stakeholders involved, and the target GHG emissions. Figure \ref{fig:ghg} highlights the detected objectives in the use of blockchain applications for GHG emissions. Additionally, we provide insights into blockchain and other supporting technologies. Table \ref{table:greenhouse-emissions} summarizes the information gathered from the surveyed literature.



\begin{table*}[thb!]
  \centering
  \caption{Blockchains applications in management of GHG emissions.}
   \renewcommand{\arraystretch}{1}

    \begin{tabular}{|>{\centering\arraybackslash}m{1.9cm}|>{\centering\arraybackslash}m{1.8cm}|>
       {\centering\arraybackslash}m{0.3cm}|>
       {\centering\arraybackslash}m{0.3cm}|>
       {\centering\arraybackslash}m{0.3cm}|>  {\centering\arraybackslash}p{0.3cm}|>
       {\centering\arraybackslash}m{1.8cm}|>
       {\centering\arraybackslash}m{2cm}|>
       {\centering\arraybackslash}m{1.8cm}|>{\centering\arraybackslash}m{1.8cm}|} \hline
       

  \multicolumn{1}{|c|}{\multirow{2}{*}{\textbf{Objectives}}}& 
  \multicolumn{1}{c|}{\multirow{2}{*}{\makecell{\bf Influencing\\ \bf Stakeholders}}} &
  \multicolumn{4}{c|}{\textbf{Blockchain Features}} & \multirow{2}{*} {\textbf{Emissions}} &
    \multirow{2}{*}{\textbf{Framework}}&
    \multirow{2}{*}{\makecell{\bf Additional\\ \bf Technologies}} & 
    \multirow{2}{*}{\makecell{\bf Challenges}}  \\ \cline{3-6}

    & 
    &{\scriptsize \bf SC}
    &{\scriptsize \bf IN}
    &{\scriptsize \bf TC}
    &{\scriptsize \bf TP}
    & & & & \\\hline
    
     Check compliance \cite{GW3,GW4,GW8, GW12} & Industry, government, companies &
      {\scriptsize{\checkmark}} & &  & {\scriptsize{\checkmark}}  &
      CO \cite{GW4}, NO$_2$, SO$_2$ \cite{GW3, GW4, GW8}, N$_2$O &
      Ethereum~\cite{GW8}, Hyperledger \cite{GW3, GW12}&
       IPFS \cite{GW4}, data compression \cite{GW8} & Scalability \\\hline

    Facilitate management \cite{GW1, GW5, GW15, GW16} & Citizens, government \cite{GW5}, gas producers \cite{GW15}  &
     & & {\scriptsize{\checkmark}}  & {\scriptsize{\checkmark}} &
    CH$_4$ \cite{GW15, GW16}, H$_2$, NH$_3$, NO$_2$, SO$_2$, NH$_3$ \cite{GW1}, PM2.5 \cite{GW5},  &
     Ethereum \cite{GW5} & 5G \cite{GW1}, Cloud and edge computing \cite{GW1}, crowdsourcing \cite{GW5}  & Scalability \cite{GW5} \\ \hline

     Utilize emission credit \cite{GW9} & Motorists, industry, polluters & 
       & {\scriptsize{\checkmark}} & & &
      NO$_2$, O$_3$, PM2.5, PM10 &
      IB-AQMS blockchain/PoW & N.A. & Smart contract conditions check 
     \\\hline    

     Collect fines \cite{GW7}  & Everyone & 
       & & & {\scriptsize{\checkmark}} & 
      NO$_2$, PM2.5 &
     N.A. & N.A. & Tracking, scalability 
     \\\hline   


Reduce cost \cite{GW6} & Everyone & 
 {\scriptsize{\checkmark}} & & & &
NO$_2$, O$_3$, PM2.5, PM10 &
Ethereum & IPFS & Cost, IPFS reliability
     \\\hline 
     
    \end{tabular}%
  \label{table:greenhouse-emissions}%
\end{table*} %

\subsubsection{Objectives of the use of Blockchain in Greenhouse Gas Emissions}
To mitigate the impact of GHG emissions, five distinct objectives have been identified: monitoring for compliance with policy limits, managing emissions by polluters, utilizing emission credits, reducing costs, and collecting fines. Works based on blockchain with these various objectives are discussed in the following.

\paragraph{Compliance check} Several studies have focused on ensuring compliance with policies, agreements, or allocated emission allowances \cite{GW1,GW2,GW3,GW8,GW12}. Han et al. \cite{GW1} proposed an architecture designed to safeguard emissions data against tampering, particularly for reports subject to compliance checks with policies or standards. Their system relies heavily on various communication technologies, including 5G, cloud computing, and edge computing, with blockchain technology utilized to preserve data integrity. Benedict et al. \cite{GW3} implemented a system for monitoring CO, CO$_2$, and smoke particulates, with the collected data secured using Hyperledger Fabric, with the final goal to secure compliance checks of transactions.
Nußbaum et al. \cite{GW8} proposed employing blockchain technology to secure data recorded by monitors tracking SO$_2$ emissions, GHG, humidity, and temperature. Their work emphasizes the importance of reducing storage requirements for securing large amounts of data.
Diniz et al. \cite{GW12} investigated the potential of blockchain technologies in enforcing compliance with the Kyoto Protocol, particularly in monitoring companies and other GHG producers. However, they also noted that stakeholders may be slow to adopt the technology or reluctant to use it.

\paragraph{Facilitate Management}

Dai et al. \cite{GW5} presented a system aimed at facilitating GHG emission management, which shares several similarities with compliance checking approaches.
Mumcular et al. \cite{GW15} proposed a blockchain-based system to monitor both renewable and non-renewable methane gas, with the goal of verifying gas quality and reducing gas consumption at the University of British Columbia.
Chen et al. \cite{GW16} investigated a methodology for quantifying methane emissions from various sources, including industrial facilities and oil wells. Their work has applications in incentivizing emission reduction, ensuring compliance, and facilitating emission trading. 
 
\paragraph{Emission Credit}
In this scenario, producers of GHG emissions may be granted credits or allowances to emit a certain amount, and blockchain could serve as a reliable mechanism for verifying compliance.
Rana et al. \cite{GW9} proposed a blockchain-based system to facilitate the use of emission credits by citizens. This approach would streamline transactions related to the management of individual emission credits, allowing for their expenditure, transfer, or exchange among individuals.

\paragraph{Fine Collection Management}
The process of fine collection necessitates an efficient payment system to monitor payments made by environmental policy offenders.
Nizeyimana et al. \cite{GW7} proposed a system for tracking fine collection and managing payments imposed on those who exceed GHG emission limits. The adoption of blockchain technology in this context ensures the implementation of a secure payment system.

\paragraph{Cost Reduction}
Sofia et al. \cite{GW6} focused their study on reducing costs in the use of blockchain for recording various GHG emissions. One strategy was to reduce the cost of storage and for that they resorted to using Inter-Planetary File System (IPFS). However, the use of IPFS also requires looking into reliability \cite{medina2024ami}. The cost of blockchain implementation is an important factor that one must consider for actual implementations and it is, therefore, a topic that deserves further study.


\subsubsection{Blockchain Features Sought in Greenhouse Gas Emissions Management}

Transparency and integrity, facilitated by the immutability property of blockchain, are key motivations for employing blockchain in monitoring and measuring GHG emissions. These properties are sought by studies on compliance verification, management facilitation, and fine collection. However, achieving transparency often requires additional functionalities to grant different stakeholders access to the recorded data. Additionally, studies on emission credits and cost reduction aim to ensure data integrity through blockchain. Furthermore, the feature of tracing is also sought by studies focused on management facilitation.

\subsubsection{Emissions}

Owning to the diversity of GHG emissions, studies have focused on subsets of emissions; no study has considered them all. For example, studies on compliance have focused on SO$_2$ and N$_2$O, but those for facilitating the management of emissions, included CH$_3$, H$_2$, SO$_2$, PH$_3$, NH$_3$. On the other hand, studies on emission credit and cost reduction have included NO$_2$, O$_3$, and also PM1, PM$2.5$ and PM10. A possible reason for such selective sets of emissions is that some emissions may be considered more prevalent in some cases or regions. Another reason is the availability of sensors; some sensors may be more available than others. Note that CO and CO$_2$ are also GHG, but those are particularly addressed in Section \ref{sec:carbon-management}.

\subsubsection{Influencing Stakeholders}

We call influencing stakeholders to those who might have access to the collected data and modify it if they experience a conflict of interest against environmental sustainability. In the applications of blockchain, the group of stakeholders who may have an interest in influencing the recorded data are industry \cite{GW1,GW3}, government for applications of compliance \cite{GW4,GW8}, everyone, including citizens and government in applications for facilitating the management of recorded data \cite{GW5}, motorists, industry, and any other generator of GHG in the management of emission credit \cite{GW9}, and everyone for applications of fine collection \cite{GW7} and cost management.

\subsubsection{Blockchain Frameworks used in Greenhouse Gas Emissions}

The most commonly reported blockchain framework used for recording GHG emissions is Ethereum \cite{GW6,GW8}. Other frameworks, such as IBM-AQMS \cite{GW9} and Hyperledger \cite{GW3}, have also been adopted. However, the reported studies did not provide information about the rationale behind their choice of framework. This lack of information suggests that the availability of blockchain frameworks may play a significant role in making the adoption of blockchain more accessible.

\subsubsection{Supporting Technologies}

Given the expansive nature of monitoring GHG emissions, crowd-sourcing approaches have been considered \cite{GW5}. Additionally, technologies such as 5G links have been utilized to transmit data directly to blockchain repositories, minimizing data exposure \cite{GW1}. Some studies have explored the use of IPFS to store recorded data in large, public off-chain databases, offering enhanced security protection \cite{GW4,GW6}. Furthermore, data compression techniques have been employed to reduce storage requirements \cite{GW8}.

\subsubsection{Challenges of Existing Blockchains}

The plethora of works focusing on recording and monitoring GHG emissions converge on scalability as a major challenge \cite{GW5,GW8}. Managing numerous sensors required for accurate evaluations of GHG generation in large urban areas poses difficulties in centralized management. One proposed approach involves outsourcing the monitoring of such emissions \cite{GW5}, although implementing a methodology to crowdsource while maintaining data immutability presents significant challenges, such as scalability. Nevertheless, this approach could extend sensing coverage to large areas. The scalability required not only affects the blockchain itself but also off-chain storage and storage in general \cite{GW6}.

Other challenges include implementing actionable smart contracts to monitor changes and compliance with emissions limits for emissions credit management \cite{GW9}. Tracking for fine collection management for heavy emission producers is another challenge, alongside scalability.

As described in this section, there are other critical emissions besides CO$_2$ that contribute to atmospheric damage. Examples of those emissions are HFCs, PFCs, SF$_6$, and NF$_3$. Studies linking multiple emissions are still needed. Moreover, GHG emissions are continuously generated, necessitating accurate monitors equipped with sensors for various emissions.

Furthermore, the reviewed literature focuses more on proposing systems that may not yet be implemented and tested. Therefore, there is an evident gap in developing systems that prioritize the implementation of blockchain and observe its performance under real testing scenarios. Having an available blockchain with flexible and robust application seems to be the most critical and immediate challenge.

\subsubsection{Other Existing Surveys on Greenhouse Gas Emissions}

In the field of management of GHG emissions, as an emerging application of blockchain, there is a notable lack of comprehensive surveys on the topic. A recent survey on the use of computer science approaches to GHG emissions was presented \cite{GW13}. This work focuses on leveraging various information systems and other computer science technologies for sustainability applications. However, the survey primarily analyzes the population of papers in the literature across different topics. Rani et al. \cite{GW14} provide a review of sustainable and energy-aware blockchain technologies and their impact on the environment. The review also briefly discusses many applications where sustainable blockchain finds utility.

In contrast to existing surveys, our study aims to uncover the unique features of blockchain technology and the specific cases that motivated its adoption for monitoring and managing GHG emissions.

\subsection{Carbon Management}
\label{sec:carbon-management}

As the primary greenhouse gas, the rising concentrations of carbon dioxide have heavily contributed to global warming and climate change \cite{solomon2009irreversible}. In response to the urgent need to radically reduce emissions, global initiatives by governments, industries, and organizations have focused on monitoring and regulating carbon emissions. These efforts include establishing carbon markets to trade carbon credits and developing technologies for capturing and storing carbon to mitigate climate change \cite{UNClimateAction,EUETS,EPAETS,DoECarbonCapture}. Carbon monitoring and trading leverage tools that follow the Kyoto Protocol \cite{KyotoProtocol}, which binds developed countries to limit and potentially reduce greenhouse gas emissions according to agreed targets. A key element of the Kyoto Protocol is that it allows countries to trade emission permits to meet their targets. This trading mechanism necessitates the monitoring of carbon emissions and the tracking of carbon trading between various stakeholders both within and across countries. Drastic measures, such as capturing carbon dioxide from the environment and storing it in depleted oil and gas reservoirs or deep saline aquifers, have also been considered \cite{wilberforce2021progress_carbonCapture}.

\begin{figure}[htp!]
    \centering
\includegraphics[width=0.8\columnwidth]{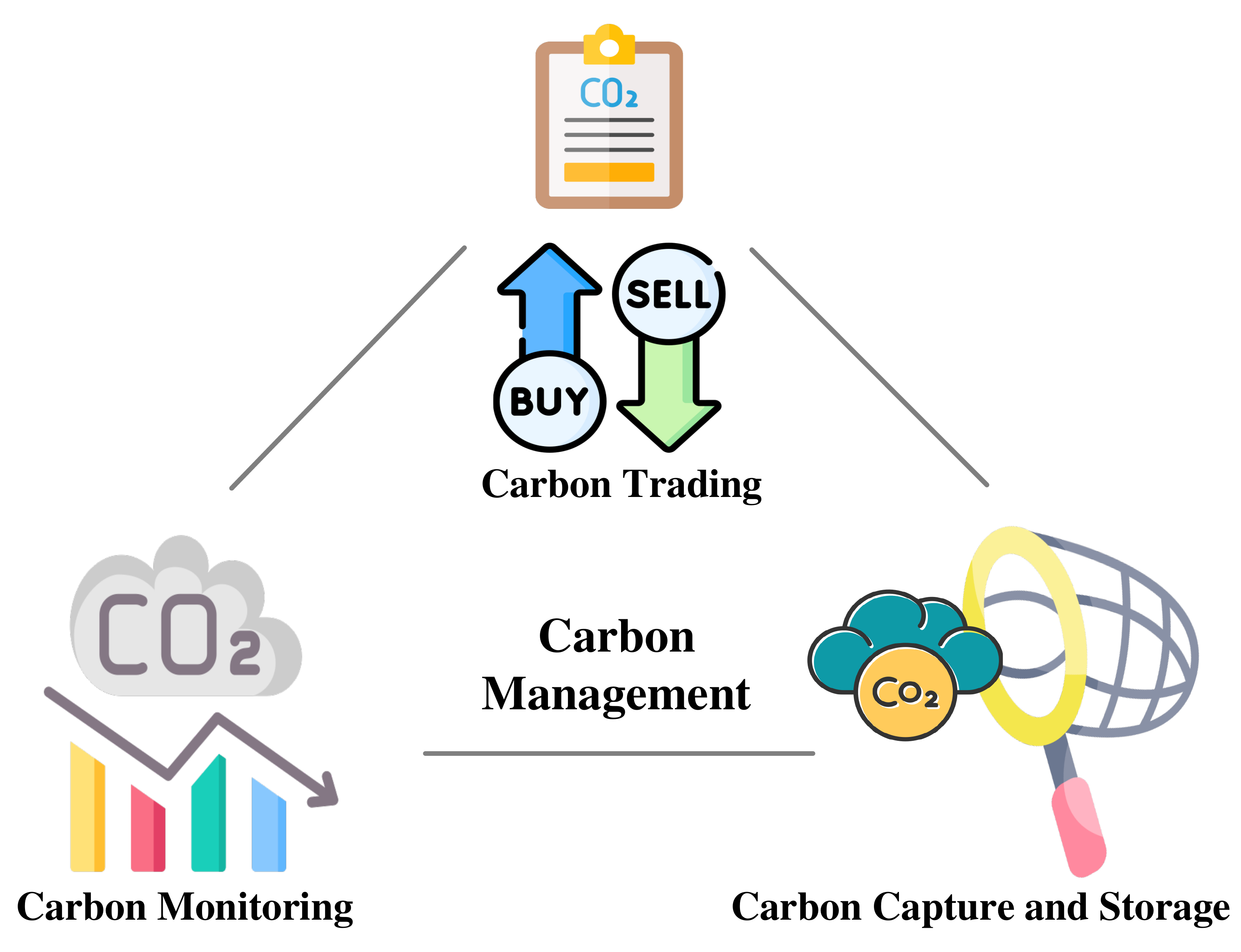}
    \caption{Three objectives pursued in blockchain applications for Carbon management.}
    ~\label{carbonManagementFigure}
\end{figure}

In 2005, the European Union (EU) implemented the first international Emission Trading System (ETS) \cite{EUETS}. The EU ETS assigns a cap on emission allowances for its participants. Those with emissions below the cap can trade their surplus carbon credits to others who cannot meet their emission targets, providing incentives for participants to reduce their emissions. A reported emission reduction of 35\% from the installations in the EU ETS demonstrates its effectiveness in curbing carbon emissions. Currently, the EU ETS encompasses the monitoring and trading of carbon emissions from electricity and heat generation, energy-intensive industry sectors, aviation, and maritime transport, with a growing number of sectors participating.

Despite showing promising results, the EU ETS is susceptible to potential single points of failure due to its centralized nature \cite{nguyen2021b_CM8,hu2020delegated_CM17}. Additionally, it lacks public trust because of the human intervention required in reporting and verifying emissions \cite{al2020hierarchical_CM7,kazicarbon_CM24}. To address these issues, several research groups have explored using the decentralized and immutable features of blockchain technology for emission monitoring and trading systems. 

A survey that reviews the application of blockchain to supply chain, IoT, energy grid management, finance, carbon trading, and big data, from the perspective of United Nations Sustainable Development Goals (SDGs), argues that blockchain can help enhance environmental sustainability~\cite{parmentola2022blockchain_CM16}. The survey discusses six examples using blockchain technology on climate change topics to show the possible positive impacts.
However, in our survey, we focus on blockchain applications for carbon management and categorize them based on their objectives.  Figure \ref{carbonManagementFigure} is a representation of these objectives. We outline the influencing stakeholders, blockchain features, blockchain framework, and consensus algorithm, adopted supporting technologies, and challenges to be addressed for these works. Table~\ref{CMTable} shows a summary of the classification of these works according to the identified objectives.  

\begin{table*}[t!]
  \caption{Blockchain-based systems for carbon management.}
   \renewcommand{\arraystretch}{1}
    \begin{tabular}{|>{\centering\arraybackslash}p{1.2cm}|>{\centering\arraybackslash}p{1.6cm}|@{}>{\centering\arraybackslash}p{3.0cm}@{}|@{}>{\centering\arraybackslash}p{0.4cm}@{}|@{}>{\centering\arraybackslash}p{0.4cm}@{}|@{}>{\centering\arraybackslash}p{0.4cm}@{}|@{}>{\centering\arraybackslash}p{0.4cm}@{}|@{}>{\centering\arraybackslash}p{0.4cm}@{}|@{}>
    {\centering\arraybackslash}p{0.4cm}@{}|@{}>{\centering\arraybackslash}p{0.4cm}@{}|@{}>
    {\centering\arraybackslash}p{0.4cm}@{}|@{}>{\centering\arraybackslash}p{3.0cm}|>{\centering\arraybackslash}p{1.8cm}|>{\centering\arraybackslash}p{1.8cm}|}\hline

    \multicolumn{1}{|c|}{\multirow{2}{*}{\textbf{Objective}}}&
    \multicolumn{1}{c|}{\multirow{2}{*}{\makecell{\bf Sector}}}&
    \multicolumn{1}{c|}{\multirow{2}{*}{\makecell{\bf Influencing\\ \bf Stakeholders}}}&
    \multicolumn{8}{c|}{\bf Blockchain Features}&
    \multirow{2}{*}{\makecell{\bf Framework\\\bf (Consensus)}}&
    \multirow{2}{*}{\makecell{\bf Supporting\\\bf Technologies}}&
    \multirow{2}{*}{\makecell{\bf Challenges}}\\\cline{4-11}

    & & & 
    \textbf{\scriptsize SC}&
    \textbf{\scriptsize DS}&
    \textbf{\scriptsize IN}&
    \textbf{\scriptsize PY}&
    \textbf{\scriptsize TC}&
    \textbf{\scriptsize TK}&
    \textbf{\scriptsize TP}&
    \textbf{\scriptsize TR}& & & \\
    \hline

    
    {}& {Building and construction \cite{zhong2022blockchain_CM6,woo2020new_CM13}}& {Contractors, sub-contractors, building owners, government, general public}& {\scriptsize \checkmark}& {}& {\scriptsize \checkmark}& {}& {\scriptsize \checkmark}& {}& {\scriptsize \checkmark}& {}& {Hyperledger Fabric (Kafka)~\cite{zhong2022blockchain_CM6}, Hyperledger Fabric~\cite{woo2020new_CM13}}& {Database~\cite{woo2020new_CM13}, sensor}& {Raw data fraud~\cite{woo2020new_CM13}}\\
    \cline{2-14}
    
    {Monitor carbon}&{Environment \cite{yan2019environmental_CM10}} &{Environmental department, general public} & {}& {}& {}& {}& {\scriptsize \checkmark}& {}& {\scriptsize \checkmark}& {}& {N/A}& {Sensor}& {Scalability}\\
    \cline{2-14}
    
    
    {}& {Food industry~\cite{shakhbulatov2019blockchain_CM26}}& {Food supply chain}& {}& {}& {}& {}& {}& {}& {}& {\scriptsize \checkmark}& {Carbon Footprint Chain (Raft-like)} & {N/A}& {N/A}\\
    \hline
    
    {}& {General public~\cite{kim2020blockchain_CM1}}& {Individuals}& {\scriptsize \checkmark}& {\scriptsize \checkmark}& {}& {\scriptsize \checkmark}& {}& {\scriptsize \checkmark}& {\scriptsize \checkmark}& {}& {Not specified (PoR)}& {AI}& {Scalability, data privacy, implementation cost}\\
    \cline{2-14}

    {}&{Corporate~\cite{effah2021carbonC_CM2,al2020hierarchical_CM7,hu2020delegated_CM17,yuan2018design_CM18,khaqqi2018incorporating_CM20,al2015bitcoin_CM21,jiang2022influencing_CM23,kazicarbon_CM24}}& {Companies, government}& {\scriptsize \checkmark}& {\scriptsize \checkmark}& {\scriptsize \checkmark}& {\scriptsize \checkmark}& {\scriptsize \checkmark}& {\scriptsize \checkmark}& {\scriptsize \checkmark}& {}& {FISCO BCOS (PBFT)~\cite{effah2021carbonC_CM2}, Multi-level blockchain~\cite{al2020hierarchical_CM7}, Hyperledger Fabric (DPoR)~\cite{hu2020delegated_CM17}, Hyperledger Fabric (Solo)~\cite{yuan2018design_CM18}, Multichain~\cite{khaqqi2018incorporating_CM20}, Bitcoin~\cite{al2015bitcoin_CM21}}& {Sensor~\cite{effah2021carbonC_CM2,al2020hierarchical_CM7}, reputation system~\cite{hu2020delegated_CM17,khaqqi2018incorporating_CM20}}& {N/A}\\
    \cline{2-14}
    
    {Trade carbon}& {Transportation \cite{lu2022stricts_CM3,eckert2020blockchain_CM4,li2021blockchain_CM5,nguyen2021b_CM8}}& {Fuel producers, vehicle manufacturers, vehicle users}& {\scriptsize \checkmark}& {\scriptsize \checkmark}& {\scriptsize \checkmark}& {\scriptsize \checkmark}& {\scriptsize \checkmark}& {\scriptsize \checkmark}& {\scriptsize \checkmark}& {}& {Hyperledger Fabric (PBFT)~\cite{lu2022stricts_CM3}, Hyperledger Iroha (YAC)~\cite{eckert2020blockchain_CM4}, Ethereum~\cite{nguyen2021b_CM8}} & {VANETs~\cite{lu2022stricts_CM3}}& {Data privacy, environmental impacts of blockchain, scalability~\cite{li2021blockchain_CM5}}\\
    \cline{2-14}
    
    {}& {Energy~\cite{muzumdar2022permissioned_CM9,hua2020blockchain_CM12}}& {Prosumers, microgrid traders}& {\scriptsize \checkmark}& {}& {\scriptsize \checkmark}& {\scriptsize \checkmark}& {\scriptsize \checkmark}& {\scriptsize \checkmark}& {\scriptsize \checkmark}& {}& {Hyperleger (Kafka)~\cite{muzumdar2022permissioned_CM9}, Ethereum~\cite{hua2020blockchain_CM12}}& {Priority and reputation system~\cite{muzumdar2022permissioned_CM9}}& {N/A}\\
    \cline{2-14}
    
    {}& {Marine ecosystem \cite{zhao2022research_CM14}}& {Blue carbon traders, government}& {\scriptsize \checkmark}& {\scriptsize \checkmark}& {}& {}& {\scriptsize \checkmark}& {\scriptsize \checkmark}& {\scriptsize \checkmark}& {}& {N/A}& {N/A}& {N/A}\\
    \cline{2-14}
    
    {}& {Building and construction \cite{woo2020blockchain_CM15}}& {Building occupants, building occupants, emission reduction assessor and verifier, and carbon credit registry}& {\scriptsize \checkmark}& {\scriptsize \checkmark}& {\scriptsize \checkmark}& {}& {\scriptsize \checkmark}& {\scriptsize \checkmark}& {\scriptsize \checkmark}& {}& {N/A}& {N/A}& {Raw data fraud}\\
    \cline{2-14}
    
    {}& {Fashion industry~\cite{fu2018blockchain_CM19}}& {Clothing manufacturers, the authority, the auditor, and the individuals}& {}& {}& {}& {}& {}& {}& {\scriptsize \checkmark}& {}& {N/A}& {N/A}& {N/A}\\
    \hline

    {Capture and store carbon}& {Industry~\cite{bachman2023incentivizing_CM25}}& {CCS plants, individuals, businesses, government}& {}& {\scriptsize \checkmark}& {\scriptsize \checkmark}& {}& {}& {\scriptsize \checkmark}& {\scriptsize \checkmark}& {}&{Not specified (PoUW)}& {Sensor}& {N/A}\\
    \hline

    \end{tabular}%
  \label{CMTable}%
\end{table*}%

\subsubsection{Objectives of the use of Blockchain in Carbon Management}

The objectives of the reviewed literature are categorized as: carbon monitoring~\cite{zhong2022blockchain_CM6,woo2020new_CM13,yan2019environmental_CM10,shakhbulatov2019blockchain_CM26}, carbon trading~\cite{kim2020blockchain_CM1,effah2021carbonC_CM2,lu2022stricts_CM3,eckert2020blockchain_CM4,li2021blockchain_CM5,al2020hierarchical_CM7,nguyen2021b_CM8,muzumdar2022permissioned_CM9,hua2020blockchain_CM12,zhao2022research_CM14,woo2020blockchain_CM15,hu2020delegated_CM17,yuan2018design_CM18,fu2018blockchain_CM19,khaqqi2018incorporating_CM20,al2015bitcoin_CM21,jiang2022influencing_CM23,kazicarbon_CM24}, and carbon capture and storage (CCS)~\cite{bachman2023incentivizing_CM25}. Carbon monitoring primarily focuses on measurement, reporting, and verification (MRV), often employing sensors that directly measure carbon dioxide emissions \cite{carbonSensor}. Carbon trading, on the other hand, not only promotes the monitoring of carbon emissions but also treats these emissions as tradable commodities, thereby motivating stakeholders to engage actively in emission reduction strategies.

\paragraph{Carbon Monitoring}

The importance of carbon monitoring extends beyond environmental agencies to include various stakeholders affected by regulatory measures on emissions and all individuals impacted by climate change-induced weather variations. The building and construction sector, known as one of the largest sources of greenhouse gases, emits significant amounts of carbon during the production, transport, and recycling of materials such as cement and steel \cite{building}. To address these challenges, Zhong et al. \cite{zhong2022blockchain_CM6} developed a blockchain application for on-site pollutant monitoring tailored to the construction industry. This system enhances data integrity and provides transparency for all project stakeholders. Similarly, Woo et al. \cite{woo2020new_CM13} introduced a blockchain-based MRV system designed to verify the energy efficiency of new buildings, increasing trust without the need for intermediaries. This system records a building's energy performance data directly on the blockchain, ensuring its accuracy and accessibility.

Yan et al. \cite{yan2019environmental_CM10} introduced a blockchain-based environmental monitoring system designed to ensure data integrity from monitoring stations and sensor networks owned by governmental or private entities. Employing a three-dimensional monitoring architecture, the system records data on air quality, water resources, soil conditions, and ecological factors, providing a comprehensive approach to environmental monitoring.

In the food supply chain domain, Shakhbulatov et al. \cite{shakhbulatov2019blockchain_CM26} proposed a Carbon Footprint Chain (CFC) to track the pre-consumption stages of the food life cycle. Utilizing a cluster-based approach, CFC enhances resilience against collusion and other threats by dynamically changing cluster leadership. Experimental results demonstrate CFC's superior performance in block creation time, leader election time, and average transaction time compared to Bitcoin. Furthermore, the blockchain implementation exhibits scalability without issues when operating with a larger number of nodes.

\paragraph{Carbon Trading}

Once carbon emission data is accurately monitored, participants in the carbon monitoring system are incentivized to engage in the trading market. In this market, participants have the opportunity to buy and sell carbon allowances to offset their emissions \cite{EUETS}. Numerous studies have explored carbon trading among individuals, companies, and various economic sectors, proposing the integration of blockchain technology to leverage its unique features.

One such proposal is an emission rights verification system designed for carbon trading, utilizing blockchain to oversee carbon credit transactions among stakeholders with diverse interests \cite{kim2020blockchain_CM1}. This governance system also incorporates AI within mobile cloud environments to authenticate trading rights and safeguard against malicious activities.

In the realm of carbon trading among companies \cite{effah2021carbonC_CM2,al2020hierarchical_CM7,hu2020delegated_CM17,yuan2018design_CM18,khaqqi2018incorporating_CM20,al2015bitcoin_CM21,jiang2022influencing_CM23,kazicarbon_CM24}, the potential for compromised emission data poses significant challenges. Such inaccuracies not only result in errors and inconsistencies in emission data but also impede decision-making and policy formulation regarding carbon emissions \cite{effah2021carbonC_CM2}. To address these concerns, a blockchain-based ETS was proposed and implemented on an enterprise-level consortium blockchain platform, a Chinese consortium blockchain, FISCO BCOS \cite{effah2021carbonC_CM2}. Similarly, Hu et al. \cite{hu2020delegated_CM17} introduced DPoR to enhance the security and efficiency of the system, where transactions are influenced by both the offer price and the reputation value of emitting enterprises. Research on the influencing factors for companies adopting blockchain technology in carbon trading reveals that the willingness of companies to embrace blockchain affects their proficiency in mastering the technology \cite{jiang2022influencing_CM23}. Additionally, Kazi et al. \cite{kazicarbon_CM24} proposed a game-theory-based ETS concept to automate carbon trading, leveraging blockchain to ensure transparency in the trading process. This approach incorporates a game-theory-based auction model to facilitate direct negotiation of carbon trading and optimize bidding strategies.

Blockchain-based solutions for carbon trading have emerged across various economic sectors, including the transportation industry, where significant opportunities exist for reducing carbon emissions \cite{lu2022stricts_CM3,eckert2020blockchain_CM4,li2021blockchain_CM5,nguyen2021b_CM8}. Lu et al. \cite{lu2022stricts_CM3} introduced STRICTs, a blockchain-enabled motor vehicle ETS, designed to address evasion issues caused by emitters falsifying carbon emission offsets. This solution utilizes smart contracts to automate audits of carbon emissions and fines for emission violations, employing Hyperledger Fabric as the underlying blockchain prototype to ensure high throughput. Similarly, Eckert et al. \cite{eckert2020blockchain_CM4} proposed cBSMD, a blockchain-based system for trading individual emissions, considering emissions from individuals traveling in a multi-modal network comprising buses and private vehicles. With low latency and high throughput capabilities for 3,186 user nodes, cBSMD offers a promising approach to carbon trading. Additionally, Nguyen et al. \cite{nguyen2021b_CM8} introduced B-ETS, a vehicle-to-vehicle ETS, demonstrating through simulations that increased vehicle connectivity can lead to reduced vehicle emissions.

In the energy sector, blockchain technology has emerged as a promising tool for tracing carbon emissions and enabling transparent trading of carbon credits \cite{muzumdar2022permissioned_CM9,hua2020blockchain_CM12}. Muzumdar et al. \cite{muzumdar2022permissioned_CM9} proposed a priority-based auction mechanism for emission trading, aiming to incentivize the use of renewable energy and deter deforestation. In this model, sellers are prioritized based on their submitted carbon credits, while buyers' priority is determined by their reputation scores, reflecting their renewable energy usage and carbon demand reduction efforts. Prosumers play a crucial role in this context, utilizing small-scale plants to generate energy from renewable sources and contributing to the local supply-demand energy balance. Hua et al. \cite{hua2020blockchain_CM12} introduced a blockchain application tailored for prosumers, facilitating P2P carbon and energy trading. Through blockchain technology, prosumers can engage in energy and carbon allowance transactions with neighboring peers, thereby achieving regional energy balance.

In marine and coastal ecosystems, the ocean carbon sink can store more carbon dioxide than terrestrial sinks, such as tropical rainforests and temperate forests \cite{blueCarbon}. This stored carbon dioxide is often referred to as blue carbon. In this context, Zhao et al. \cite{zhao2022research_CM14} introduced a conceptual blockchain framework aimed at supporting blue carbon trading, with the goal of unlocking the vast potential of the ocean carbon sink.

The building sector is one of the largest contributors to greenhouse gas emissions. To address this challenge, Woo et al. \cite{woo2020blockchain_CM15} proposed a theoretical framework for blockchain-based carbon trading. This framework simplifies the process of acquiring carbon credits using smart contracts. Building owners initiate the credit acquisition process by submitting preliminary plans, which are then assessed and verified by predefined smart contracts on the blockchain. Through this trading framework, building owners are incentivized by the potential profit earned from selling credits.

The fashion apparel industry, known for its resource-intensive manufacturing processes, including significant water and energy consumption, stands to gain from participation in the blockchain-based carbon trading market \cite{fu2018blockchain_CM19}. Fu et al. showed that the implementation of a trading framework could effectively reduce carbon emissions associated with apparel manufacturing processes.

\paragraph{Carbon Capture and Storage}

CCS contains three major steps: capture, transportation, and storage. The stakeholders involved in the CCS ecosystem include industries, government, certification agencies, and the general public.
Bachman et al.~\cite{bachman2023incentivizing_CM25} introduced a new native token on a blockchain that uses 
PoUW to incentivize carbon removal. The CCS facilities are rewarded with tokens for the removal and storing of carbon dioxide. With this incentive mechanism, CCS facilities compete with each other for the amount of captured and stored carbon emissions. 

\subsubsection{Influencing Stakeholders}

The influencing stakeholders in blockchain-based carbon management systems include individuals, the government, and parties that directly participate in the process of carbon monitoring, credit trading, or CCS. In carbon monitoring, stakeholders span various sectors including building and construction, environmental agencies, municipalities, and the food industry. These stakeholders may include building contractors and owners, environmental departments, local or regional governments, participants in the food supply chain, and the general public. Carbon trading involves stakeholders from diverse sectors such as corporate entities, transportation, energy, marine ecosystems, building and construction, and the fashion industry, alongside the general public and government bodies. In CCS, influencing stakeholders encompass CCS plants, businesses emitting carbon dioxide, individuals, and governmental bodies.

\subsubsection{Blockchain Features Sought in Carbon Management}

In carbon monitoring, transparency, traceability, and tracking features of blockchain are enabled to ensure stakeholders have access to a transparent platform that provides traceable records of carbon emissions and tracks carbon footprints. Smart contracts can also be employed to perform automatic carbon emission compliance checking and distribute various incentives to participants. In carbon trading, smart contracts can automate the trading between participants and facilitates the transfer of their payments. Incentives can also be distributed to participants for carbon emission reductions in the form of tokens or carbon credits. Data security, transparency, and traceability are other common features enabled by blockchain-based emission trading systems. In carbon capture and storage (CCS) initiatives, blockchain technology leverages features such as data security, transparency, incentives, and tokenization to encourage CCS facilities to actively remove carbon dioxide from the environment.

\subsubsection{Blockchain Framework used in Carbon Management}

To implement blockchain for carbon management, frameworks such as Hyperledger Fabric~\cite{zhong2022blockchain_CM6,woo2020new_CM13,hu2020delegated_CM17,yuan2018design_CM18,lu2022stricts_CM3,muzumdar2022permissioned_CM9}, Ethereum~\cite{nguyen2021b_CM8,hua2020blockchain_CM12}, Carbon Footprint Chain (CFC)~\cite{shakhbulatov2019blockchain_CM26}, FISCO BCOS~\cite{effah2021carbonC_CM2}, Multichain~\cite{khaqqi2018incorporating_CM20}, Bitcoin~\cite{al2015bitcoin_CM21}, and Hyperledger Iroha~\cite{eckert2020blockchain_CM4} have been implemented.

Unfortunately, the work in CCS have not identified blockchain frameworks that could be used for this purpose. However, the proposed conceptual framework that incentivizes CCS facilities uses PoUW~\cite{bachman2023incentivizing_CM25}. The {\it useful work} in this consensus scheme is estimated in terms of the amount of carbon dioxide the CCS facilities capture and store. 

For carbon monitoring and carbon trading, depending on the application, a public blockchain such as Ethereum~\cite{nguyen2021b_CM8,hua2020blockchain_CM12} and Bitcon~\cite{al2015bitcoin_CM21}, or a private blockchain, such as Hyperledger Fabric~\cite{zhong2022blockchain_CM6,woo2020new_CM13,hu2020delegated_CM17,yuan2018design_CM18,lu2022stricts_CM3,muzumdar2022permissioned_CM9} have been reported used. Moreover, a public blockchain might be used when individuals are involved in the carbon monitoring or trading system~\cite{nguyen2021b_CM8,hua2020blockchain_CM12,al2015bitcoin_CM21}.

Hyperledger Fabric, a private and permissioned blockchain, is more energy-efficient compared to public blockchain systems \cite{zhong2022blockchain_CM6}. Various consensus algorithms can be implemented on Hyperledger Fabric. Some examples are Kafka \cite{zhong2022blockchain_CM6,muzumdar2022permissioned_CM9}, PBFT \cite{effah2021carbonC_CM2,lu2022stricts_CM3}, DPoR \cite{hu2020delegated_CM17}, and Solo \cite{yuan2018design_CM18}, where the implementation consists of one ordering node. On CFC, a raft-like consensus is implemented to achieve high performance and scalability \cite{shakhbulatov2019blockchain_CM26}. FISCO BCOS is an open-source enterprise-level financial blockchain commonly used by companies in China \cite{effah2021carbonC_CM2}. A PBFT consensus is used on FISCO BCOS. 

Multichain is an open blockchain platform \cite{khaqqi2018incorporating_CM20}. One of the features of multichain is to restrict the actions that a node could perform. Hyperledger Iroha can be used to implement a private and a public blockchain \cite{eckert2020blockchain_CM4}. This framework adopts YAC as consensus algorithm.

\subsubsection{Supporting Technologies}

Various supporting technologies are proposed to be used in combination with blockchain for carbon monitoring and trading. For carbon monitoring, a database is often used to store the data obtained from sensors and IoTs to achieve fast access time and low cost per record stored \cite{woo2020new_CM13}. By storing data in both a blockchain and a database, one can employ the immutability feature of blockchain as well as the fast and cost-effective features of the database.

For carbon trading, Sadawi et al.~\cite{al2020hierarchical_CM7} proposed to incorporate the Blockchain of Things (BoT) paradigm where sensors are connected to the consortium blockchain using specific protocols. This approach reduces human intervention and ensures data immutability. 
To govern carbon trading system, Kim and Huh~\cite{kim2020blockchain_CM1} proposed to use AI to detect carbon emissions anomalies. In the transport sector, to enable trading between vehicles, VANETs are employed~\cite{lu2022stricts_CM3}.

To fight against climate change with carbon trading, a reputation system is often employed along with a carbon trading scheme to incentivize the emitting enterprises to reduce their carbon emissions~\cite{hu2020delegated_CM17,khaqqi2018incorporating_CM20,muzumdar2022permissioned_CM9}. This system rewards the participants with a higher reputation when they put a sizeable effort into avoiding deforestation and using renewable energy. The rewards give participants advantages in the carbon trading market.

\subsubsection{Challenges of Existing Blockchains}

While blockchain-based carbon management systems hold significant promise for reducing carbon emissions across various sectors, they are not without challenges. These systems may encounter issues such as data privacy, scalability, cost, and the risk of raw data fraud during the emission data collection process. The inherent transparency of blockchain technology ensures symmetrical information sharing among stakeholders, but it also raises concerns regarding the handling of confidential details belonging to companies or individuals. In the context of CCS, scalability poses a significant challenge due to the high computational and energy costs associated with adding blocks to the blockchain network. The scalability issue becomes more pronounced as more participants or sensors are integrated into the carbon management system, potentially impacting its performance. Additionally, the energy consumption of blockchain technology raises concerns about its carbon footprint and its implications for carbon emissions. In carbon trading systems, the process of collecting emission data may be vulnerable to raw data fraud, particularly when human intervention is involved. While blockchain is often touted for its ability to bring trust to systems, the susceptibility of human participation to data manipulation poses a challenge to this claim~\cite{woo2020blockchain_CM15}.


\subsection{Solid Waste Management}
\label{sec:solid-waste}

Solid waste is the disposed materials that are generated from our daily activities, movable, not directly used, and permanently discarded \cite{nanda2021municipal}.
Solid waste management process may be regulated based on its type and property and by different entities.

{\it Municipal waste} 
comprises biodegradable and non-biodegradable components from households, offices, small-scale institutions, and commercial enterprises and is usually processed by the waste management department at the municipality level~\cite{nanda2021municipal}. 

{\it Electronic waste (e-waste)} includes a wide range of end-of-life electric and electronic equipment that may be handled by electronic recycling solution providers.

{\it Medical waste} generated from hospitals, research institutions, health care teaching institutes, clinics, and laboratories are plastics that might contain biohazards that need to be handled by specialized medical waste solution providers~\cite{SWM18}.

{\it Agricultural waste} is the residue from growing and processing raw agricultural products, including crop residues, livestock waste, and agro-industrial and aquaculture waste that needs to be properly handled on the farms~\cite{SWM04}.


{\it Hazardous waste} 
are heavy metals, asbestos, or toxic chemicals from various industries. When the volume of hazardous waste is higher than safety standards, it could pose a threat to public health and the environment. Thus, it must be properly handled by specialized staff and processes~\cite{lagrega2010hazardous}.

Independent of their type, most of the solid waste ends up in landfills and yet, it significantly impacts the environment and human health by contaminating the soil and groundwater, and by generating methane and carbon dioxide during decomposition~\cite{el1997environmental}. Solid waste management (SWM) aims to reduce waste volume and minimize its impact on public health and the environment. 

\begin{figure}[htp!]
    \centering
\includegraphics[width=0.95\columnwidth]{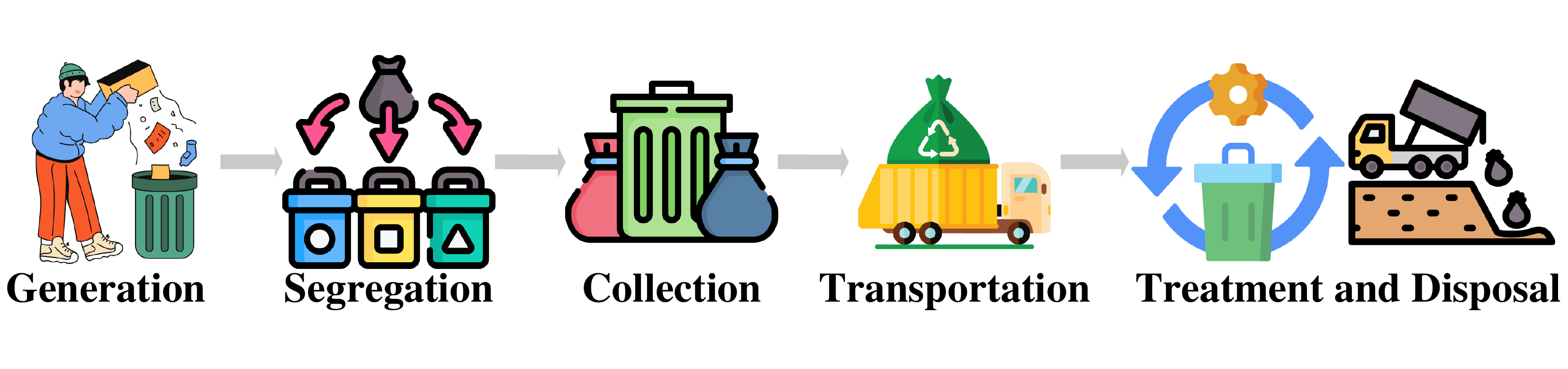} 
    \caption{Stages of solid waste management for the application of blockchain.}
    ~\label{phasesSWM}
\end{figure}

Effective solid waste management typically encompasses the following phases: generation, segregation, collection, transportation, treatment, and disposal in an environmentally friendly manner, as shown in Figure~\ref{phasesSWM}. 
{\it Waste Generation} represents all the waste from households, offices, institutions, commercial enterprises, municipal services, medical, agricultural, and industrial facilities~\cite{nanda2021municipal}. {\it Waste Segregation} refers to sorting waste according to its type~\cite{SWM14}. {\it Waste Collection} gathers solid wastes from their point of generation~\cite{SWM06}. {\it Waste Transportation} is the process of transporting the collected waste to treatment or disposal facilities. {\it Waste Treatment and Disposal} is the process where solid waste is either treated, repurposed, or disposed of in landfills.

Waste management faces numerous challenges, primarily due to the difficulty of tracking every piece of solid waste and heavy reliance on manual data recording. This dependence brings consequentially various issues, such as recording errors and data security concerns. Unfavorable consumer behavior, inaccurate waste segregation, insufficient waste disposal facilities, inadequate recycling, and low information security exacerbate these problems~\cite{SWM05}. Information asymmetry arises when regulators and regulated institutions lack equal access to information. 
Addressing these interconnected issues requires a holistic and comprehensive approach to improve waste management practices effectively. Blockchain has been explored as a solution for different stages of solid waste management.

\subsubsection{Objectives of the use of Blockchain in Solid Waste Management}

The objectives of existing blockchain-based systems for SWM include incentivizing waste segregation and collection, monitoring waste transportation, and improving SWM system.  We categorize them based on these objectives as shown in Table \ref{tab:swm} along with the type of waste, the reported influencing stakeholders, blockchain features, adopted framework and consensus, supporting technologies, and the remaining challenges to be addressed.   

\paragraph{Incentivize Waste Segregation}

One of the challenges that solid waste management faces is the lack of effective systems for segregating waste materials. Recyclable waste materials can generate revenue as they can be repurposed, but that requires an effective system to handle that. Paturi et al.~\cite{SWM14} proposed a municipal waste management system that uses blockchain and IoT to distribute rewards for individuals who properly dispose of waste. This approach aims at encouraging individuals to sort the waste according to the category and then discard it in designated smart bins equipped with IoT technology. Individuals who properly discard solid waste receive cryptocurrencies as rewards.

\paragraph{Incentivize Waste Collection}

Proper disposal is important to reduce the possible harm that end-of-life e-waste brings to the environment. E-waste management oversees the process of environmentally friendly e-waste disposal. The first step is e-waste collection. Blockchain has been considered for incentivizing e-waste collection~\cite{SWM02,SWM08}. Customers who discard their end-of-life e-waste at the retailer or e-waste center get rewards through smart contracts~\cite{SWM02, SWM08}. In addition, if the retailer or e-waste center fails to collect the e-waste, a penalty is imposed on the retailer or e-waste center. For agricultural waste, Zhang et al.~\cite{SWM04} proposed a blockchain model to incentivize farmers to collect and trade their waste, such as crop straw and animal residue at enterprises that convert the waste to clean energy or agricultural by-products. By collecting and trading the waste with enterprises, farmers can get tokens that can be traded with other farmers for energy or fertilizer. A blockchain system can also address the challenges of limited citizen participation and company cooperation in municipal waste collection~\cite{SWM06,SWM21}. 
França et al.~\cite{SWM06} proposed to use cryptocurrency, Green Coin, to replace printed cards used to incentivize individuals to participate in waste collection. The cryptocurrency can also be used to trade products in local registered stores.

\paragraph{Monitor Waste Transportation}
To monitor hazardous waste transportation and reduce crimes in hazardous waste transfer, such as illegal dumping and falsification of hazardous waste quantities, Song et al.~\cite{SWM23} proposed to use blockchain to mitigate information asymmetry during hazardous waste transportation. This approach allows real-time monitoring of hazardous waste transfer and tracing of all information during the transfer process. For municipal waste transportation, Saad et al.~\cite{SWM27} proposed an IoT-driven blockchain-based system using VANETs to enable real-time monitoring of waste transportation and waste vehicle performance.

\paragraph{Improve Solid Waste Management}
To improve SWM, Gopalakrishnan et al.~\cite{SWM12} proposed a blockchain system to improve waste management by enhancing product tracking and encouraging responsible waste management practices. The system also incorporates a customer reward program, which motivates households and end-users to share information and encourages them to bring recyclable waste to nearby facilities for proper disposal. The authors also conducted a cost analysis of the blockchain-based solid waste management traceability system considering both the cost of implementing the system as well as the cost for customers and proposed an optimization model to help municipalities understand and improve the effectiveness of their waste management operations~\cite{SWM07}. 
To address the challenges in the medical supply chain and waste disposal tracking during COVID-19 pandemic, two blockchain systems were proposed to enhance the efficiency, transparency, and security of the supply chain and waste management process~\cite{SWM18,SWM24}. Ahmad et al.~\cite{SWM18} proposed a blockchain, based on Ethereum, that uses IPFS as decentralized storage to securely manage data related to COVID-19 medical equipment supply chain and waste management. 
The authors implemented four smart contracts that handle order management, lot and ownership management, waste shipment, and registration to log data along the supply chain and waste management. 
Le et al.~\cite{SWM24} proposed Medical-Waste Chain, a waste collection, classification, and treatment management system on Hyperledger Fabric to automate the waste management processes for medical equipment and supplies after usage. The system's goals include minimizing medical waste generation, facilitating real-time recording and sharing of waste treatment data, ensuring transparency among stakeholders, preventing information forgery, and promoting the reuse and efficient scheduling of waste materials.

\begin{table*}[thb!]
  \centering
  \caption{Blockchain applications for solid waste management.}
   \renewcommand{\arraystretch}{1}
    \begin{tabular}{|@{}>
    {\centering\arraybackslash}p{1.9cm}@{}|@{}>{\centering\arraybackslash}m{2.4cm}@{}|@{}>{\centering\arraybackslash}p{2.8cm}@{}|@{}>{\centering\arraybackslash}p{0.5cm}@{}|@{}>{\centering\arraybackslash}p{0.5cm}@{}|@{}>{\centering\arraybackslash}p{0.5cm}@{}|@{}>{\centering\arraybackslash}p{0.5cm}@{}|@{}>{\centering\arraybackslash}p{0.5cm}@{}|@{}>{\centering\arraybackslash}p{0.5cm}@{}|@{}>{\centering\arraybackslash}p{0.5cm}@{}|@{}>{\centering\arraybackslash}p{1.8cm}@{}|@{}>{\centering\arraybackslash}p{2.5cm}@{}|@{}>{\centering\arraybackslash}p{3cm}@{}|}\hline

    \multicolumn{1}{|c|}{\multirow{2}{*}{\textbf{Objective}}}&
    \multicolumn{1}{c|}{\multirow{2}{*}{\makecell{\bf Waste Type}}}&
    \multicolumn{1}{c|}{\multirow{2}{*}{\makecell{\bf Influencing \\ \bf Stakeholders}}}&
    \multicolumn{7}{c|}{\textbf{Blockchain Features}}&
    \multirow{2}{*}{\makecell{\bf Framework\\ \bf (Consensus)}}&
    \multirow{2}{*}{\makecell{\bf Supporting\\ \bf Technologies}}&
    \multirow{2}{*}{\makecell{\bf Challenges}}\\\cline{4-10}

    & & &
    \textbf{\scriptsize SC}&
    \textbf{\scriptsize DG}&
    \textbf{\scriptsize DS}&
    \textbf{\scriptsize IN}&
    \textbf{\scriptsize TC}&
    \textbf{\scriptsize TP}&
    \textbf{\scriptsize TR}&
    &
    &
    \\
    \hline

    \multirow{4}{*}{}\makecell{\\Incentivize \\waste \\segregation\\~\cite{SWM14}} & 
    \multirow{4}{*}{}\makecell{Municipal} & 
    \multirow{4}{*}{}\makecell{Government, \\corporations, \\and individuals} & 
    {\scriptsize \checkmark}&
    {}&
    {}&
    {\scriptsize \checkmark}& 
    {\scriptsize \checkmark} &
    {\scriptsize \checkmark} &
    &
    {Ethereum} & {IoT, QR code}& {N/A}\\
    \hline

     \multirow{4}{*}{}\makecell{} & 
     \multirow{2}{*}{\makecell{Electronic\\~\cite{SWM02,SWM08}}}& 
     \multirow{4}{*}{}\makecell{Manufacturer, \\supplier, retailer, \\customers,\\ and e-waste center}& 
      {\scriptsize \checkmark}& {}& {}& {\scriptsize \checkmark}& {}&{\scriptsize \checkmark} & {\scriptsize \checkmark} & Ethereum & {IoT, sensors, barcodes, browser extensions}& {Incentivation \cite{SWM02}, Security \cite{SWM08}}\\
    \cline{2-13}

    \multirow{3}{*}{}\makecell{Incentivize \\waste \\collection\\\cite{SWM04,SWM06}\\ \cite{SWM02,SWM08,SWM21}}& 
    \multirow{3}{*}{}\makecell{Agricultural~\cite{SWM04}}& 
    \multirow{3}{*}{}\makecell{Farmers, \\waste-to-energy \\plant, \\government agency}& 
    {}& {}& {}& {}& {}&{\scriptsize \checkmark} & {} & {N/A}& {QR Code, IoT}& {Scalability}\\
    \cline{2-13}
    
    \multirow{4}{*}{\makecell{}} & 
    \multirow{4}{*}{}\makecell{\\Municipal\\~\cite{SWM06,SWM21}}& 
    {Everyone}& 
     {\scriptsize \checkmark}& {}& {}& {\scriptsize \checkmark}& {}&{} & {\scriptsize \checkmark} & \makecell{Ethereum\\~\cite{SWM06}}& {Fog computing, GPS, RFID~\cite{SWM21}, IoT~\cite{SWM21}}& 
    {Implementation and tokenization~\cite{SWM06,SWM21}}\\
    \hline

   \multirow{5}{*}{\makecell{Monitor \\waste \\transportation\\~\cite{SWM23,SWM27}}}& 
   \multirow{2}{*}{\makecell{Hazardous~\cite{SWM23}}}& 
   \multirow{3}{*}{}\makecell{Environmental \\protection agency, \\hazardous waste\\ producers, \\transportation \\and treatment\\ facilities}& 
   {\scriptsize \checkmark}& {}& {}& {}& {\scriptsize \checkmark}&{\scriptsize \checkmark} & {\scriptsize \checkmark} & {N/A}& 
   \multirow{5}{*}{}\makecell{IoT, GPS}& 
   {Validate input data, data ownership, lack of regulation}\\
    \cline{2-13}

    \multirow{3}{*}{}\makecell{} & 
    \multirow{3}{*}{}\makecell{Municipal~\cite{SWM27}}& 
    \multirow{3}{*}{}\makecell{Waste vehicles\\ company}& 
    {}& {}& {\scriptsize \checkmark}& {}& {}&{\scriptsize \checkmark} & {\scriptsize \checkmark} & {N/A}& 
    \multirow{3}{*}{}\makecell{UHF, VANETs, \\IoT, Geo-fencing\\ techniques}& {N/A}\\
    \hline

   \multirow{6}{*}{\makecell{\\Improve \\solid waste\\management\\~\cite{SWM05,SWM18}\\ \cite{SWM07,SWM12,SWM17,SWM24}}}& 
   \multirow{4}{*}{}\makecell{Municipal\\~\cite{SWM05,SWM07}\\ \cite{SWM12,SWM17}}
   & 
   \multirow{4}{*}{}\makecell{Sorting, recycling, \\incinerator, \\landfill facility, \\and customers}& 
   {\scriptsize \checkmark}& {\scriptsize \checkmark}& {}& {\scriptsize \checkmark}& {\scriptsize \checkmark}&{\scriptsize \checkmark} & {\scriptsize \checkmark} &
   \multirow{4}{*}{}\makecell{Ethereum \\(PoW)~\cite{SWM05},\\ Ethereum\\~\cite{SWM12,SWM17}}& 
   \multirow{4}{*}{}\makecell{QR code~\cite{SWM12},\\ IoT~\cite{SWM17}}& 
    \multirow{4}{*}{}\makecell{Resilience in\\ adopting blockchain\\ technology~\cite{SWM07}, \\scalability~\cite{SWM12}}\\
    \cline{2-13}

     \multirow{4}{*}{}\makecell{}& 
     \multirow{4}{*}{}\makecell{Medical\\~\cite{SWM18,SWM24}}& 
     \multirow{4}{*}{}\makecell{Medical center,\\ transportation, \\recycling plants\\ sorting facilities \\and waste centers}& 
     {\scriptsize \checkmark}& {}& {\scriptsize \checkmark}& {}& {\scriptsize \checkmark}&{\scriptsize \checkmark} & {\scriptsize \checkmark} & 
      \multirow{4}{*}{}\makecell{Ethereum~\cite{SWM18},\\ Hyperledger \\Fabric~\cite{SWM24}}& {IPFS~\cite{SWM18}}& 
    \multirow{4}{*}{}\makecell{Data privacy, \\lack of regulation, \\smart contract\\ vulnerabilities, \\scalability~\cite{SWM18}, \\data validation~\cite{SWM24}}\\
    \hline

    \end{tabular}%
  \label{tab:swm}%
\end{table*}%

\subsubsection{Influencing Stakeholders}
Influencing stakeholders in solid waste management include government agencies and companies that handle municipal waste and everyone who produces waste \cite{SWM14,SWM06,SWM21,SWM27,SWM05,SWM07,SWM12,SWM17}. Blockchain is used to incentivize segregation and collection and improve the efficiency of operation. For special types of waste, such as e-waste \cite{SWM02,SWM08}, agriculture \cite{SWM04}, hazardous \cite{SWM23}, and medical \cite{SWM18,SWM24}, special care needs to be taken by the transportation, treatment, and disposal facilities to ensure proper government regulations are followed and avoid illegal waste disposal that harms the environment.  

\subsubsection{Blockchain Features Sought in Solid Waste Management}

Smart contracts, transparency, and tracking are the most sought-after blockchain features to address the challenges of information asymmetry in solid waste management. Data governance is reported for improving municipal SWM to control access to the record. Data security is adopted by monitoring municipal waste transportation and improving biomedical waste management. Incentivization is adopted for waste segregation and collection as well as improving municipal waste management to provide incentives for the participation of desired waste handling behaviors. Traceability is also reported in approaches for incentivizing waste segregation, monitoring hazardous waste transportation, and improving municipal waste management.   

\subsubsection{Blockchain Frameworks used in Solid Waste Management}

Several blockchain systems have been proposed, but many examples are reported as possible approaches. Only a few provide comprehensive insights into the blockchain framework and the consensus mechanisms suitable for their intended applications. Nevertheless, the available information on these frameworks remains limited. Among the surveyed papers, Ethereum~\cite{SWM02,SWM05,SWM06,SWM08,SWM12,SWM14,SWM17,SWM18} and Hyperledger Fabric~\cite{SWM24} emerge as the frameworks in use. The reported consensus algorithm includes PoW~\cite{SWM05}.

\subsubsection{Supporting Technologies}

Various technologies, such as IoT~\cite{SWM04,SWM14,SWM17,SWM21,SWM23}, GPS~\cite{SWM21}, QR codes~\cite{SWM04,SWM12,SWM14}, and RFID~\cite{SWM21},
are crucial in collecting data throughout the waste sorting, collection, transportation, and treatment and disposal process. They are often integrated into smart waste bins or waste trucks. A decentralized storage system, IPFS is also used for off-chain data logging \cite{SWM18}. 
Ultra-high frequency (UHF) and VANETs are used to efficiently retrieve sensor data over extended distances. VANETs are used to detect the location of waste bins, while UHF readers and tags are used to identify waste bins~\cite{SWM27}. 
Geo-fencing techniques are employed for effective waste monitoring and timely collection from dump spots~\cite{SWM27}.



Various computing applications have been explored to address challenges in waste management systems, such as managing IoT devices, hosting services, and applications. Fog computing is proposed to handle these tasks with reduced latency and faster processing by storing the blockchain ledger~\cite{SWM21}. Cloud computing is proposed to store data and providing high availability and access~\cite{SWM06}. Edge Computing is a potential application that can be explored to minimize the local transaction time~\cite{SWM04}. VANETs enable real-time location sharing and asset transportation management~\cite{SWM27}.

\subsubsection{Challenges of Existing Blockchains }







Some challenges commonly identified with blockchain technology are also identified in blockchain-based SWM systems. They are the implementation and monetary value determination of the cryptocurrency~\cite{SWM06,SWM21}, data sharing boundary~\cite{SWM18,SWM23}, scalability~\cite{SWM04,SWM12}, lack of regulation on blockchain technology~\cite{SWM23}, resilience in adopting blockchain technology~\cite{SWM07}, and smart contract vulnerabilities~\cite{SWM18}. Data standards and how to validate data from off-chain storage to blockchain are needed to ensure interoperability among the systems~\cite{SWM23,SWM24}.

\subsubsection{Existing Surveys on Blockchain-based Solid Waste Management}

\begin{table*}[htbp]
    \centering
    \caption{Existing surveys on blockchain applications for solid waste management.}~\label{SWM_surveys}
    \begin{tabular}{|c|>{\raggedright\arraybackslash}m{8.5cm}|c|c|c|}\hline
          {\bf Survey} & \multicolumn{1}{c|}{\bf Objective} & {\makecell{\bf No. of Surveyed\\ \bf Papers}} & {\bf Year}\\\hline
          \cite{9540705}& Explore how blockchain enhances waste management in smart cities &116&2021\\\hline
          \cite{dondjio2021blockchain}& {Study how blockchain improves waste management through enhanced public awareness, transparency, and stakeholder trust} &63& 2021\\\hline
          \cite{bamakan2022towards}& Investigate how blockchain can be used in managing hospital waste  & 207&2022\\\hline
          \cite{baralla2023waste}&{Assess blockchain waste management benefits for the circular economy and identify adoption drawbacks in the waste sector} &98&2023\\\hline
          \cite{jiang2023blockchain}& Adoption of blockchain in waste management and challenges and opportunities when combined with AI and IoT  &180&2023\\\hline
          \cite{singh2023systematic}&Explores the opportunities of using blockchain, IoT, GPS, and machine learning in waste management&184&2023\\\hline
    \end{tabular}
\end{table*}

Several surveys have explored waste management leveraged by blockchain~\cite{9540705, dondjio2021blockchain, bamakan2022towards, baralla2023waste,jiang2023blockchain, singh2023systematic}. We outlined their focus, the number of publications reviewed and the year it was published as shown in Table~\ref{SWM_surveys}. These surveys show an increasing interest in this topic from academia. 

Reviews on how combining blockchain with IoT~\cite{9540705, jiang2023blockchain} and machine learning models can improve SWM efficiency~\cite{jiang2023blockchain} have been explored with benefit and challenges analysis. 
Bamakan et al.~\cite{bamakan2022towards} present how blockchain can enhance the management of medical waste. Baralla et al.~\cite{baralla2023waste} and Jiang et al. \cite{jiang2023blockchain} analyze various approaches on using blockchain in circular economy for SWM. Different from these surveys, our survey provides an updated detailed category of blockchain-based solutions for SWM and identifies the motivation for using blockchain for the management of different types of waste, as some of this waste presents challenges for reuse and recycling. 

\subsection{Plastic Management}
\label{sec:plastic-waste}

Plastics represent a major source of pollution to the environment, with the potential to affect the plants, animals, the complete food chain, and human health \cite{macleod2021global,chae2018current,wilcox2015threat}. Their robust scope of applications made it be considered "the material of thousand uses," and that has motivated indiscriminate production and consumption of it \cite{plastic_bakelite}. Plastic is a key contributor to the municipal solid waste. However, because of its vast impact on the environment, we dedicate to it its own section.

\begin{figure}[htp!]
    \centering
\includegraphics[width=0.9\columnwidth]{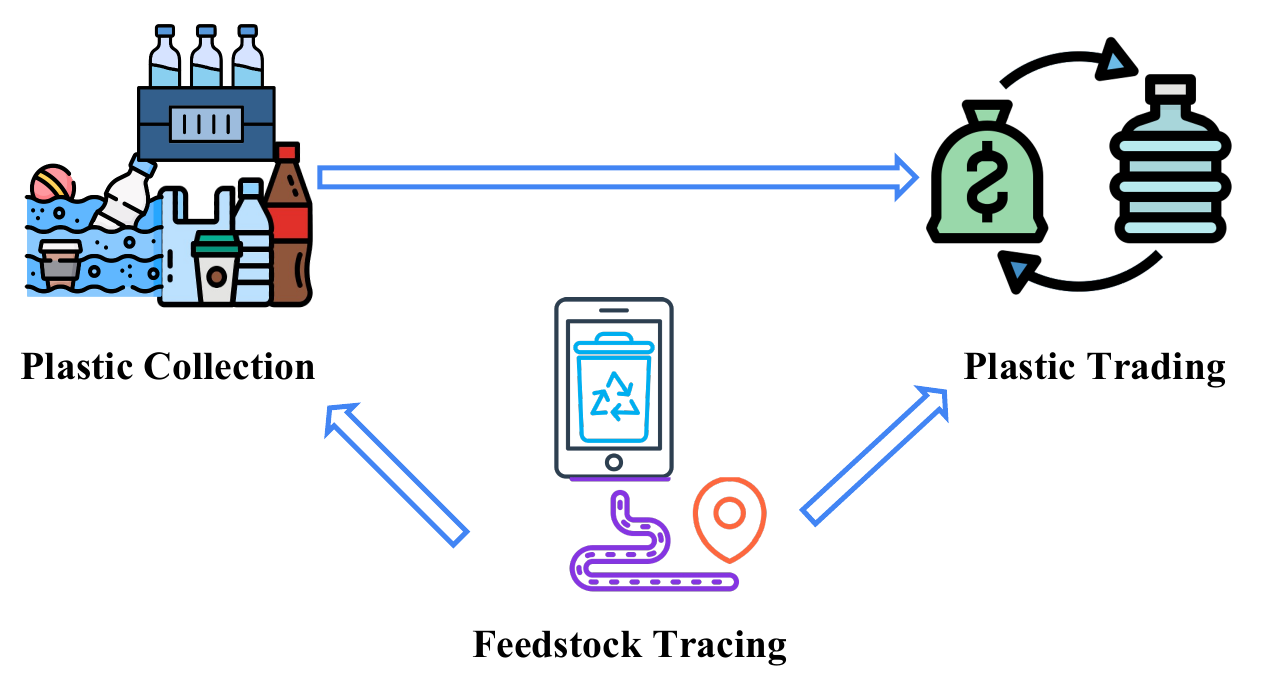}
    \caption{Main objectives sought in the management of plastic with existing blockchain-based systems.}
    ~\label{fig:plastic}
\end{figure}

Around 400 million metric tons (MMT) of plastic waste is presently produced \cite{plastic_un}. Independently of its origin, plastic waste eventually ends up in the ocean, being carried through rivers, industrial, and storm-water runoffs. 
In the ocean, plastic disintegrates into microplastic particles ($<$5~mm in size), impacting the ocean wildlife and its ecosystem \cite{plastics_rochman}, and in turn affects the seafood and the people who consume it.

The {\it throwaway living} style, based on single-use plastics, has been globally embraced, and that greatly exacerbates the pollution by plastic \cite{plastics_LAtimes}. Recent studies show that the projected plastic production will grow to 1480 MMTs by 2050 \cite{plastic_atlas}.
The large and global production of synthetic plastics has been creating two major environmental problems: greenhouse gas emissions and pollution.

Municipalities deal with plastic through incineration, disposal to landfills, recycling, or exporting to other countries/regions.  
Some of these options are partially or not beneficial for the environment in the long term because their main goal is rather to reduce the presence of local plastic waste. For example, incineration generates greenhouse gases. Plastic deposited in landfills takes a long time to degrade. 
Local recycling of plastic is considered more sustainable for the environment, elevating the promise of reducing not only waste but also the manufacturing of more products from recycled plastics. However, recycling is expensive and their quality is considered inferior to that of virgin plastics~\cite{plastics_recycleQuality}. Exporting plastic waste overseas reduces its local presence but whether it reduces pollution and its carbon footprint depends on its final processing. 
It is reported that about 14\% of plastic waste is incinerated, 40\% is disposed of in landfills, 14\% is recycled, and the remaining 32\% is mismanaged, spread in the environment, and eventually ending in rivers and oceans~\cite{plastic_atlas}.

To incentivize plastic recycling, trace the feedstock of plastics, and support plastic trading, blockchain has been explored as a potential solution to provide trust and traceability, and generate digital credit to enhance plastic management. 
Table \ref{tab:plastic-recycling} shows the reported blockchain-based solutions for plastic management categorized based on their objectives.

\subsubsection{Objectives of the use of Blockchain in Plastic Waste }

As shown in Table \ref{tab:plastic-recycling}, the objectives of blockchain-based approaches for plastic-waste management include: incentivizing plastic collection, plastic feedstock tracing, and plastic trading. Figure \ref{fig:plastic} shows these objectives. 

\paragraph{Incentivize plastic collection}
\label{sec:plasitc_bottle}

Beverage packaging using polyethylene terephthalate (PET) plastic is one of the major sources of plastic pollution because of the large global consumption of bottled beverage drinks. It is reported that \mbox{Coca-Cola} alone produces \mbox{108~billion} plastic bottles annually \cite{plastics_cocacola}. 
There is a myriad of issues that exacerbate plastic pollution with plastic bottles. However, the leading one is the lack of liability from all stakeholders. Therefore, blockchain has been proposed as an approach to ensure that single-use PET bottles are properly managed. 
Wankm{\"u}ller et al. \cite{plastics_PR3} developed a process model to study different tokenization incentives for plastic collection. They also examined how these incentives can support participation in a circular plastic supply chain. The pilot study of blockchain tokenization showed that the high level of operator convenience for plastic collection is an important factor in helping accelerate large-scale system deployment.  
The study adopted MultiChain, a private blockchain, as its blockchain platform. 

Both residential and commercial areas dispose of various classes of recyclable plastics. 
In 2013, a Canadian for-profit initiative launched a blockchain solution, Plastic Bank to unify localized plastic-waste collections for a more effective and global recycling \cite{plastics_PR8}. \mbox{Plastic Bank} uses IBM's Hyperledger Fabric and enables traceability to incentivize marginalized waste collectors, alleviate economic hardship in developing countries, and reduce global pollution. 
In Germany, a similar initiative, Deposy, developed specialized vending machines connected to IOTA, a blockchain-like distributed ledger and cryptocurrency infrastructure, to financially incentivize the recovery of digitally tagged urban plastics from consumers \cite{plastics_PR15}. The approach assists in cost-efficient and effective recycling through sorting plastic waste. 

Lynch et al. \cite{plastics_PR17littercoin} proposed a cloud-based web interface, named OpenLitterMap, that classifies, records, and visualizes geospatial locations of plastic waste using crowdsourced data. The approach is based on reports generated from the geotagged images of plastic waste. The images are shared by participants and garbage collectors are informed of the location of the reported plastic waste. The approach uses littercoin, an Ethereum-based cryptocurrency, to reward contributors for sharing verified images. Mondal et al. \cite{PR20_mondal2022blockchain} proposed a conceptual framework that uses Hyperledger Besu (Ethereum Virtual Machine, EVM) to incentivize plastic recycling by connecting users to plastic collection companies. Here blockchain is not used to handle transactions but as a medium to connect stakeholders.

\begin{table*}[t!]
  \centering
  \caption{Blockchain applications for plastic waste management.}
   \renewcommand{\arraystretch}{1}
    \begin{tabular}{|@{}>{\centering\arraybackslash}p{2cm}@{}|@{}>{\centering\arraybackslash}p{2.5cm}@{}|@{}>{\centering\arraybackslash}p{2.3cm}@{}|@{}>{\centering\arraybackslash}p{0.6cm}@{}|@{}>{\centering\arraybackslash}p{0.6cm}@{}|@{}>{\centering\arraybackslash}p{0.6cm}@{}|@{}>{\centering\arraybackslash}p{0.6cm}@{}|@{}>{\centering\arraybackslash}p{0.6cm}@{}|@{}>{\centering\arraybackslash}p{3cm}@{}|@{}>{\centering\arraybackslash}p{3cm}@{}|@{}>{\centering\arraybackslash}p{1.7cm}@{}|}\hline

    \multicolumn{1}{|c|}{\multirow{2}{*}{\bf Objective}}&
    \multicolumn{1}{c|}{\multirow{2}{*}{\makecell{\bf Category}}}&
    \multicolumn{1}{c|}{\multirow{2}{*}{\makecell{\bf Influencing\\\bf Stakeholders}}}&
    \multicolumn{5}{c|}{\bf Blockchain Features}&
    \multirow{2}{*}{\makecell{\bf Framework\\\bf (Consensus)}}&
    \multirow{2}{*}{\makecell{\bf Supporting\\\bf Technologies}}&
    \multirow{2}{*}{\makecell{\bf Challenges}}\\\cline{4-8}

    & & &
    \textbf{\scriptsize SC}&
    \textbf{\scriptsize IN}&
    \textbf{\scriptsize TC}&
    \textbf{\scriptsize TP}&
    \textbf{\scriptsize TR}& & &\\
    \hline
    \multirow{10}{*}{\makecell{Incentivize \\plastic \\ collection \\\cite{PR20_mondal2022blockchain}\\\cite{plastics_PR3,plastics_PR8,plastics_PR15,plastics_PR17littercoin}}}& 
    \multirow{2}{*}{Bottles \cite{plastics_PR3}} &
    Bottling companies, consumers & 
    & 
    \multirow{2}{*}{\scriptsize\checkmark} & 
    & 
    & 
    &
    Multichain/Fine-grain permission & 
    \makecell{Digital badge, \\mobile app}&
    \multirow{10}{*}{\makecell{Data \\security\\ \cite{plastics_PR15}}}\\
    \cline{2-10}
    & 
    \multirow{3}{*}{\makecell{Urban \\ \cite{plastics_PR8,plastics_PR15,plastics_PR17littercoin,PR20_mondal2022blockchain}}} &
    Manufacturers, consumers, collection agencies & 
    \multirow{3}{*}{\scriptsize\checkmark} & 
    \multirow{3}{*}{\scriptsize\checkmark} & 
    & 
    & 
    &
    Hybrid blockchain \cite{plastics_PR8}, Hyberledger Besu \cite{PR20_mondal2022blockchain}, IOTA (Fast probabilistic consensus) \cite{plastics_PR15}& 
    Digital badge, mobile app, IoT devices \cite{plastics_PR8,plastics_PR15,PR20_mondal2022blockchain}&
    \\
    \cline{2-10}
    & 
    \multirow{3}{*}{\makecell{Coastal  and \\ packaging \\\cite{plastics_PR17littercoin}}} &
    Waste image submitting participants & 
    & 
    \multirow{3}{*}{\scriptsize\checkmark }& 
    & 
    & 
    &
    Ethereum & 
    Mobile app, crowd-sourcing geo-tagged image, relational database, cloud computing &
    \\
    \cline{2-10}
    
    \hline 
    
    \multirow{6}{*}{\makecell{Trace plastic \\feedstock \\\cite{plastics_PR7,plastics_PR16,PR10,plastics_PR12,plastics_PR14}}}&
    \multirow{3}{*}{Ocean  \cite{PR10} }&
    Collection agencies, manufacturers, certification agencies &
    &
    &
    \multirow{3}{*}{\scriptsize \checkmark} &
    &
    &
    \multirow{3}{*}{N/A} &
    \multirow{3}{*}{N/A} &
    \\
\cline{2-10}
    &
    \multirow{3}{*}{\makecell{Recycled  \\\cite{plastics_PR7,plastics_PR12},\\\cite{plastics_PR14}}}&
    Collection agencies, manufacturers, government&
    \multirow{3}{*}{\scriptsize \checkmark}&
    \multirow{3}{*}{\scriptsize \checkmark}&
    \multirow{3}{*}{\scriptsize \checkmark}&
    \multirow{3}{*}{\scriptsize \checkmark}&
    &
    Consortium blockchain \cite{plastics_PR14}&
    Digital badge \cite{plastics_PR12}, IoT devices, AI \cite{plastics_PR7} &
    \\
    \cline{2-10}
    
    \hline
    
    \multirow{5}{*}{\makecell{Support \\plastic \\trading \\\cite{plastics_PR9,plastics_PR11,PR21_alnuaimi2023blockchain,PR24_mondal2022incentivization}}}&
    \multirow{3}{*}{\makecell{Products\\ \cite{plastics_PR9,plastics_PR11}}}&
    Manufacturers, government, NGOs, consumers &
    \multirow{3}{*}{\scriptsize \checkmark} &
    \multirow{3}{*}{\scriptsize \checkmark} &
    \multirow{3}{*}{\scriptsize \checkmark} &
    \multirow{3}{*}{\scriptsize \checkmark} &
    \multirow{3}{*}{\scriptsize \checkmark} &
    
    Consortium blockchain \cite{plastics_PR11}, Hyperledger Fabric + Ethereum (PBFT + PoW) \cite{plastics_PR9}
    & N/A
    &
    \multirow{5}{*}{\makecell{Scalability\\ \cite{plastics_PR9}\\ \cite{plastics_PR11}}}\\
    \cline{2-10}
    &
    \makecell{Plastic waste \\~\cite{PR21_alnuaimi2023blockchain,PR24_mondal2022incentivization}}&
    Consumers, recycling organizations&
    \multirow{2}{*}{\scriptsize \checkmark} &
    \multirow{2}{*}{\scriptsize \checkmark} &
    \multirow{2}{*}{\scriptsize \checkmark} &
    \multirow{2}{*}{\scriptsize \checkmark} &
    \multirow{2}{*}{\scriptsize \checkmark} &
   \makecell{ Ethereum \cite{PR21_alnuaimi2023blockchain}, \\ EVM \cite{PR24_mondal2022incentivization}}&
    DApp &
    \\
    \hline
    \end{tabular}%
  \label{tab:plastic-recycling}%
\end{table*}%

\paragraph{Plastic Feedstock Tracing}
\label{sec:plastic_tracking}

Virgin plastics are cheaper to produce than recycled plastics, and therefore, they are the largest portion of fabricated plastics in the market. Regulating the production scale of virgin plastic may start with monitoring plastic feedstock in the supply chain.  Moreover, the production of virgin plastic may use a small portion of plastic from recycling. That plastic is also a portion of the feedstock that must be monitored \cite{plastics_EUreport}.

Microplastics ($>$5~mm in size) on the ocean surface have recently claimed significant attention. While plastic bottles and other plastics are of concern, 94\% of ocean plastic weight \cite{plastics_macro} comprises discarded fishing nets and lines, mostly made of PET or High-density polyethylene (HDPE) \cite{plastics_fishnets}. Monetary incentives based on plastic weight motivate the recovery of this plastic waste from the ocean. However, this incentivization can also be abused by reporting collected waste not from the ocean. As a response to this concern, a Norwegian project developed a blockchain-based solution to record and certify collections of offshore plastic waste \cite{PR10}. The project uses blockchain to record information on the clean-up time and place, plastic properties, and to issue digital certificates. The purpose is to guarantee the buyers are helping to get rid of plastic in the ocean. Trust and traceability are the main features the project claims to provide.  

A data-fusion-assisted process for the separation of plastics, leveraged by blockchain, was recently proposed \cite{plastics_PR7}. 
The approach comprises a repository and smart contracts to automate the availability and procurement of recycled feedstocks for relevant stakeholders, e.g., product manufacturers.

Dell Technologies Inc. launched a cloud-based solution to source and trace recycled plastic feedstock for establishing a sustainable supply chain \cite{plastics_PR16}. The goal of this approach is to produce digitally tagged packaged goods on recycled HDPE plastics. This solution uses AI for plastic-waste classification; it is currently deployed on the \mbox{VMWare-Blockchain} platform, which runs smart contracts for different supply-chain operations, involving collection agencies, government, transportation agencies, and others. Like Plastic Bank, this solution promotes the economic development of marginalized waste collectors through transparent payments.  

PlasticChain, a blockchain-based approach was proposed to manage data about highly recyclable raw plastics and plastic products available in the consumer market \cite{plastics_PR9}. PlasticChain comprises two blockchain platforms, one blockchain, implemented on Hyperledger Fabric, tracks information about raw material/product design and manufacturing contracts. The other blockchain, implemented on Ethereum, records data about plastic quality and producer company names for transparency. 

ReciChain, a consortium blockchain, was proposed and piloted in Brazil. It tracks plastic waste throughout the value chain through digital twins and uses blockchain to incentivize participants~\cite{plastics_PR14}. 

Waste2Wear  \cite{plastics_PR12} uses smart contracts in a blockchain to encourage the adoption of textile products manufactured from recycled PET and Polypropylene (PP) feedstock collected from land and coastal regions for a circular economy model. The blockchain-based solution is expected to facilitate transparency in the supply chain by generating traceable data about recycled plastics in digitally tagged products.

\paragraph{Plastic trading}
\label{sec:plastic_allowance}
A plastic credit-driven system that uses a recyclability index (RI) and plastic credit was proposed to support self-regulation and trading among industry participants through blockchain-enabled smart contracts \cite{plastics_PR11}. The system consists of a consortium blockchain, M-InfoChain, for RI negotiation and trading, and a public blockchain, CreditChain, to support manufacturers to publish their RI indexes to the public for auditing and transparency.  

Alnumaimi et al. \cite{PR21_alnuaimi2023blockchain} proposed an Ethereum blockchain framework with four types of smart contracts and a DApp to improve plastic collection efficiency and fairly reward participants. The smart contract implementation is available on GitHub. 

Mondal et al. \cite{PR24_mondal2022incentivization} also proposed a Blockchain-based incentivization model to encourage organizations to collect post-consumer plastics and track their supply. The system contains a main-chain and a carbon-credit chain to provide a global view of the entire supply chain. It uses EVM with smart contracts implemented on Solidity, and Truffle suite for library and smart contracts testing.

\subsubsection{Influencing Stakeholders}

The identified influencing stakeholders in blockchain applications for plastic management are bottling companies and consumers for plastic bottle collection. In urban plastic collection, the influencing stakeholders include collection agencies, waste processors, product manufacturers, and consumers \cite{plastics_PR8,plastics_PR15}. 

In plastic feedstock tracing, the influencing stakeholders include supply-chain participants, e.g, collection agencies, waste processors and product manufacturers \cite{plastics_PR7,plastics_PR16,plastics_PR14,plastics_PR12}, transportation agencies \cite{plastics_PR16}, polymer producers \cite{plastics_PR9,plastics_PR14}, feedstock suppliers \cite{plastics_PR14}, and regulatory body, government \cite{plastics_PR16}, among others.
In the collection of ocean waste, the influencing stakeholders are the waste collection agencies, waste processors, product
manufacturers, and certification agencies~\cite{plastics_PR17littercoin}. 

In plastic trading, top-level supply-chain participants, e.g., polymer producers and product manufacturers along with authorized auditors, e.g., governments and non-profit organizations (NGOs), as well as consumers are identified as influencing stakeholders. 

\subsubsection{Blockchain Features Sought in Plastic Waste}
Incentivization is the most sought-after blockchain feature for plastic management. Smart contracts were considered for urban plastic collection \cite{plastics_PR12}. 
Traceability was exclusively used for feedstock tracing and trading. Transparency was reported for tracing recycled plastics and supporting plastic trading to provide the source information of the plastics~\cite{plastics_PR12,plastics_PR11, PR24_mondal2022incentivization}. Tracking was only used to support plastic trading to match consumers with the plastic recycling organization~\cite{PR20_mondal2022blockchain, PR21_alnuaimi2023blockchain}. 

\subsubsection{Blockchain Framework used in Plastic Waste}

The majority of the blockchain solutions for plastic management adopt a multi-chain approach that uses a combination of multiple blockchains to handle credit transferring and recording tracing \cite{plastics_PR9,plastics_PR11, PR24_mondal2022incentivization}, IOTA~\cite{plastics_PR15} or a consortium blockchain \cite{plastics_PR14}. Hyperledger fabric and Ethereum are the most adopted frameworks. 
The most popular consensus algorithm is PBFT, while PoW and PoS are also considered in some of the system implementations. Many of the piloted systems by commercial organizations use proprietary consensus in their systems.  

\subsubsection{Supporting Technologies:}
Digital badges, assets, and wallets are commonly used as a credit system to incentivize plastic collection and trading. Database, image processing, smartphone apps, cloud computing, and IoT devices are often adopted in sorting, tracking plastic waste, and providing information to improve system efficiency. 

\subsubsection{Challenges of Existing Blockchains: } 
\label{sec:bc_limitations_plastics}

A few of the pilot blockchain solutions for plastic management have exposed several issues including data security \cite{iotaBreach}, scalability of the system because of the selected consensus algorithm \cite{consensus_eval}, privacy concerns~\cite{gdpr}, and energy efficiency of the blockchain framework.

\subsubsection{Existing Surveys on Plastic Waste}
\label{sec:plastic_survey}

Existing surveys on blockchain for plastic recycling explore social aspects, plastic feedstock, tracking, and case studies. 
Steenmans et al.~\cite{plastics_survey21} reviewed 21 commercial solutions (two of which are on plastic management) using first-hand interviews and secondary data.  
The reported blockchain use for waste management includes payment, recycling and reuse rewards, monitoring and tracking of waste, and smart contracts.

Bhubalan et al.~\cite{PR19_bhubalan2022leveraging} reviewed work on molecular tagging to track plastic.  
Verma et al. \cite{PR22_verma2022blockchain} focused on recycling plastics and plastic waste using AI and blockchain technology. It discusses various plastic regulation policies and AI utilization for plastic recycling, an overview of the blockchain, and its classification for waste management or plastic recycling. 

Two case studies on ongoing projects using blockchain and sensing technologies in Finland were presented by Sankaran \cite{PR23_sankaran2019carbon}: one transforms industrial carbon emissions into green fuels and one helps in efficient and sustainable segregation and recycling of plastic waste using multi-sensor-driven AI and blockchain tools. 

Rumetshofer et al. \cite{PR27_rumetshofer2023information} presented a survey on plastic markers, digital product passports, and certification systems and conducted strengths, weaknesses, opportunities, and threats analysis of these areas. 

\subsection{Food Waste Management}
\label{sec:food-waste}

Food waste encompasses not only the discarded food itself but also the energy, water, and resources expended during its production, transportation, and packaging, significantly amplifying the carbon footprint of the food supply chain \cite{FW_6}. According to the United Nations (UN) and Food Agriculture Organization (FAO), food loss occurs due to mishandling or malfunctions in production and supply systems, including instances where food is removed from the supply chain despite being suitable for human consumption \cite{FW_4}.

Food loss primarily occurs during production, post-harvest, and processing stages, while food waste predominantly occurs at the retail and consumer levels \cite{food_loss_waste}. Factors such as inadequate storage facilities, lack of cold chains, and inefficient marketing systems contribute to food loss, whereas food waste often stems from poor stock management or neglect. For the purposes of this paper, the term 'food waste' encompasses both food loss and food waste occurring throughout the supply chain.

Food waste is a significant contributor to GHG emissions and environmental pollution \cite{FW_5}. It accounts for approximately 6-8\% of all human-caused GHG emissions, equivalent to the emissions from approximately 32.6 million cars in the U.S. alone. When food waste ends up in landfills, it decomposes anaerobically, producing methane \cite{FW_5}.

The scale of food waste is staggering, with approximately one-third of all food produced globally (around 1.3 billion tons) being lost or wasted annually, as reported by the World Food Program (WFP) \cite{FW_2}. This waste occurs even as some countries and regions face famine and food insecurity simultaneously, according to FAO \cite{FW_3}.

\begin{figure}[htp!]
    \centering
\includegraphics[width=0.8\columnwidth]{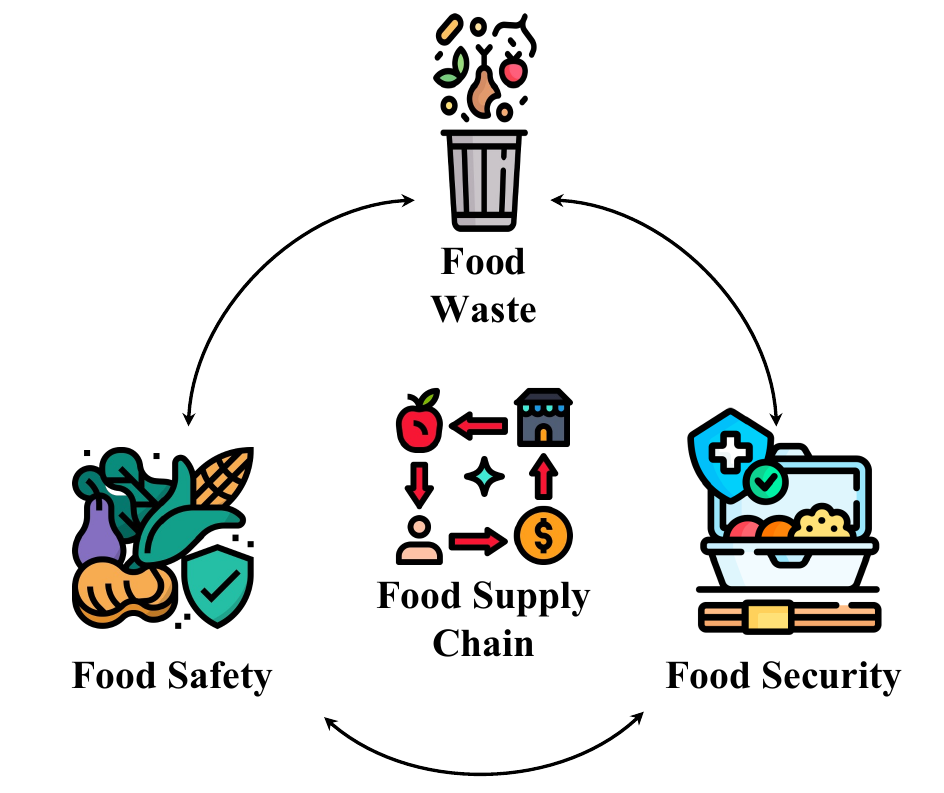}
    \caption{Four main objectives sought in existing blockchain applications for food management.}
    ~\label{fig:food}
\end{figure}

In response to these challenges, recent research has increasingly turned to blockchain technology to enhance the efficiency of food supply chains, reduce food waste, and ensure food safety and security. In this section, we categorize existing literature on blockchain proposals and applications for food waste management based on their objectives, as illustrated in Figure \ref{fig:food}. The classification of these proposals according to their identified objectives is presented in Table \ref{tab:fw}.  

\subsubsection{Objectives of the use of Blockchain in Food Waste Management}

The objectives of using blockchain in the food supply chain are to a) reduce food waste, b) ensure food safety and security, and c) improve the efficiency of the food supply chain. These objectives, shown in Figure \ref{fig:food}, are discussed in the following.  

\paragraph{Reduce Food Waste}
Various blockchain-based solutions have been proposed to address food waste, leveraging data across the supply chain to establish traceable and reliable systems for product tracking and origin tracing when necessary.

Marin et al. \cite{FW3} introduced a mobile web platform powered by Hyperledger Fabric, enabling dairy farm data tracing. Transactions are conducted online using virtual currency, with all peers maintaining ledgers for token collection. Utilizing QR codes and RFID tags, the system tracks product movement in a milk industry case study. In the event of product contamination, the system swiftly identifies the farm of origin, facilitating targeted removal of affected batches and minimizing food waste. This application aims to prevent food fraud, enhance transparency in dairy product provenance, expedite food contamination source identification, and ultimately reduce waste.

Dey et al. \cite{FW6} proposed SmartNoshWaste, a blockchain-based framework comprising two layers: the Data System Architecture and the ML Module. The Data System Architecture layer employs blockchain, QR codes, and cloud computing to digitalize and store food data, while the ML Module layer leverages this data to estimate food consumption and mitigate waste. Analyzing data across the food supply chain, SmartNoshWaste has the potential to reduce food waste at the household level by aiding individuals in managing their food consumption.

Baralla et al. \cite{FW38} introduced a method for tracing European food supply chain data utilizing a consortium blockchain, specifically Hyperledger Sawtooth, in conjunction with QR codes. This system allows the consumer to access the product history from the origin to verify product health compliance and quality.

\begin{table*}[t!]
  \centering
  \caption{Blockchain applications for food waste management.}
   \renewcommand{\arraystretch}{1}
    
       \begin{tabular}{|>{\centering\arraybackslash}m{2cm}|>
       {\centering\arraybackslash}m{1.8cm}|>{\centering\arraybackslash}m{0.2cm}|>{\centering\arraybackslash}m{0.2cm}|>{\centering\arraybackslash}m{0.2cm}|>{\centering\arraybackslash}m{0.2cm}|>{\centering\arraybackslash}m{0.2cm}|>{\centering\arraybackslash}m{4cm}|>
       {\centering\arraybackslash}m{2.4cm}|>{\centering\arraybackslash}m{1.8cm}|}
    \hline

    \multicolumn{1}{|c|}{
    \multirow{2}{*}{\textbf{Objective}}}& 
    \multicolumn{1}{c|}{\multirow{2}{*}{\makecell{\bf Influencing\\ \bf Stakeholders}}} &\multicolumn{5}{c|}{\bf Blockchain Features} &\multirow{2}{*}{\bf Framework}&\multirow{2}{*} {\makecell{\bf Supporting\\ \bf Technologies}}&\multirow{2}{*}{\makecell{\bf Challenges}} \\\cline{3-7}

    & 
    &\textbf{\scriptsize SC}
    &\textbf{\scriptsize DS}
    &\textbf{\scriptsize TC}
    &\textbf{\scriptsize TP}
    &\textbf{\scriptsize TR}& & & \\\hline

    {Reduce food waste~\cite{FW3, FW6, FW9,FW38, FW12}} &
    {Consumers, manufacturers}&
    {\scriptsize{\checkmark}}& {\checkmark}& {\checkmark} & {\checkmark}& {} 
    &{Hyperledger Fabric~\cite{FW3}, SmartNoshWaste~\cite{FW6},
    Hyperledger Sawtooth~\cite{FW38}, originChain~\cite{FW12}}

    &{ QR code~\cite{FW3,FW6, FW38}, RFID~\cite{FW3}, Cloud Computing~\cite{FW6}   }
    &{Scalability}\\\hline
   
    {Ensure food safety~\cite{FW24,FW23,FW14,FW15,FW36,FW4}} &
    {Logistics, manufacturers}&
    {\scriptsize{}}& {}& {\checkmark}& {\checkmark}& {\checkmark}
    &{\multirow{6}{*}{}\makecell{Trusted Trade Blockchain\\ Network Cloud Platform \\(TTBNCP)~\cite{FW36}, \\Not specified~\cite{FW14,FW15} \\IBM Hyperledger \\Framework~\cite{FW4,FW23},\\ \cite{FW24}}}
    &{IoT~\cite{FW36}, RFID~\cite{FW15}}
    &{Scalability, data security}\\\hline
    
    {Ensure food security~\cite{FW43,FW44,FW45,FW7}} &
    {Consumers}&
    {\scriptsize{}}& {}& {\checkmark}&  {\checkmark}& {\checkmark}
    & {Not specified} 
    &{IoT~\cite{FW45}, RFID~\cite{FW43}, ML~\cite{FW44}, PDS~\cite{FW7}}
    &{Scalability}\\\hline
    
    {Improve food supply chain~\cite{FW37,FW35,FW11,FW41,FW42}} &
    {Consumers, logistics, manufacturers}&
    {\scriptsize {\checkmark}}& {}& { {\checkmark}}& { {\checkmark}}& {}
    &\multirow{2}{*}{}\makecell{Modified blockchain~\cite{FW37}, \\ (PBFT)~\cite{FW41}}
    &{IoT~\cite{FW37}, RFID~\cite{FW37}, MAS~\cite{FW11}}
    &{Scalability, transparency}\\\hline

    \end{tabular}%
  \label{tab:fw}%
\end{table*}%

\paragraph{Ensure Food Safety}

Food safety comprises the conditions and practices that preserve food quality, and prevent contamination and the generation of food-borne illnesses. Companies and academia have explored using blockchain to improve food safety within the supply chain. In a practical example, Walmart~\cite{FW4,FW23,FW24} has reportedly used blockchain to ensure food safety among its clients. On the academic side, scientific research continues to explore alternative solutions to further improve food safety \cite{FW14,FW15,FW36}. Blockchain is mainly used for traceability and transparency properties. Food contamination and selling counterfeit or spoiled products represent a risk worldwide. The use of a blockchain-based system to track credentials for food safety checks might significantly reduce food safety concerns \cite{FW4}. 

As a growing concern worldwide, food fraud represents a supplier's action to knowingly deceive customers about the quality and contents of the foods~\cite{FW4}. Deterring food fraud requires traceability and authenticity checks along the supply chain. 
Walmart has partnered with IBM to customize two proof of concept (POC) consensus algorithms that focus on two elements of the blockchain solutions: traceability and authenticity to not only reduce food waste but also to ensure food safety~\cite{FW23, FW24}. 

Tse et al.~\cite{FW14} proposed a system that helps the government track, monitor, and audit the food supply chain and helps manufacturers record authentic transactions while eliminating the fragmented management system in China’s food supply chain. 
Food safety is achieved by combining RFID and blockchain for the entire agri-food supply chain. The system gathers and works with authentic data in the supply chain's production, processing, warehouse, distribution, and selling phases~\cite{FW15}. 


Lin et al. proposed an ecological food traceability system based on blockchain and IoT~\cite{FW36}. The system involves all parties of a smart agriculture ecosystem. With the use of IoT devices, manual recording and verification is replaced to reduce human interaction in supply chain phases, such as production, process, logistics, and storage. A smart contract is designed to help the law executor find irregularity within the supply chain on time.

\paragraph{Ensure Food Security}

Food security is the condition where people have physical and economic access to sufficient, safe, and nutritious food that meets their dietary needs or restrictions and food preferences for a healthy life \cite{FW_1}. The UN promotes food security, also known as food quality, as a solution to achieving zero hunger worldwide in their sustainable development goals \cite{FW7, FW45}. 

Green food supply chain management (GFSCM) and interrelationship modeling (ISM) techniques were implemented to reduce waste and improve food quality and safety~\cite{FW45}. Kumar et al.~\cite{FW45} proposed the use of IoT and blockchain to improve food security while reducing food waste. Chandan et al.~\cite{FW7} compared blockchain-based systems from a food security perspective while preserving SDGs. A blockchain-based public distributed system (PDS) was proposed to reduce food waste and ensure zero hunger while addressing the issues of food quality, safety, and provenance~\cite{FW7}. These proposed systems improve transparency and provide real-time stock data. This decentralized and transparent system helps an impartially and timely distribution of food to the end consumer.

Yadav et al.~\cite{FW44} used ML and blockchain to ensure food security in India. In their proposed system, 14 drivers were interviewed to collect their reponses. Classification methods were applied to the collected data to present food-related information to the main stakeholders: governmental agencies and NGOs. RFIDs were used to monitor the products within the supply chain. 
A combination of unique identifications and changeable identifications was proposed to address the food security issue~\cite{FW43}.

\paragraph{Improve Food Supply Chain}

There is an increasing interest in using blockchain to improve the food supply chain while reducing waste and improving food provenance~\cite{FW11}, food safety~\cite{FW42}, and system scalability~\cite{FW41}. 
The following works use blockchain to evaluate the current systems, identify inefficiency and flaws, and propose solutions to improve the supply chain. 

Casado-vara et al.~\cite{FW11} proposed a blockchain multi-agent system (MAS) to address the lack of information regarding food provenance. The proposed system can track shipments, authenticate origin and destinations, and provide proof that all transactions are stored and unaltered. A circular economy is an approach that can address food safety concerns and perform quality assurance. Pakseres et al. \cite{FW42} showed how combining circular economy and blockchain could improve data usage, making the supply chain more efficient, enhancing eco-efficiency, and improving traceability. 
In a more targeted approach, Liu et al.~\cite{FW41} proposed a modified PBFT consensus scheme based on clustering data to optimize existing applications that discuss food traceability using blockchain. This consensus scheme reduces resolution and communication time. 

Mondal et al.~\cite{FW37} proposed an RFID and proof-of-object-based authentication protocol, similar to PoW, to improve transparency in the food supply chain. RFID provides a unique ID to the products and helps monitor real-time quality. The system is customizable, and moisture, light, or volatile sensors can be integrated with the RFID tag.

\subsubsection{Blockchain Features Sought in Food Waste Management}

Many works motivated to enable transparency, traceability, and provenance are resorting to the use of blockchain. Consumer demand for information on the provenance and the processes performed on a product as it passes through the supply chain is increasing.

Transparency and traceability promote confidence across the processes by allowing customers to check data in the system~\cite{FW12, FW3}.  
Transparency captures data beyond traceability, providing information about product production. For example, Walmart aims at allowing food system participants to optimize supply chains and reduce food waste while encouraging accountability and incentivizing stakeholders in the food system to do the right things by safeguarding food safety every time~\cite{FW4}. 
The system ensures a close connection between all stakeholders in the supply chain and direct communication between farmer and consumer, minimizing food waste due to mislabeling and miscommunication~\cite{FW38}. 

Ensuring food safety is a global concern. Food contamination can occur and cause serious diseases or even death to small and large groups of people. Traceability~\cite{FW36, FW41, FW23, FW24, FW14, FW15} and transparency~\cite{FW23, FW24} can help solve food safety issues.

Traceability and transparency are important factors that also impact food security. They allow stakeholders to communicate better with customers, increase efficiency, and reduce risks and costs of collection in case of product recall~\cite{FW7, FW45}. Data security is used to protect the information passed to the blockchain against attacks in the early stages of the supply chain~\cite{FW43}.  

\subsubsection{Influencing Stakeholders}

Consumers are influencing stakeholders in reducing food waste and ensuring food security. It is reported that people waste about 80 million tons of food per year, which equals 149 billion meals in the same time frame~\cite{FW_6}. Food manufacturers are also critical players. Moreover, developing efficient and effective logistics is essential in reducing food loss, ensuring food safety, and improving the supply chain. Although lots of food gets wasted on the consumer side, food loss remains a major problem in the supply chain.

\subsubsection{Blockchain Frameworks used in Food Waste Management}

To reduce food waste, blockchain has been considered to ensure food safety and to provide data governance to the supply chain. For these purposes, there are academic proposals~\cite{FW36, FW14, FW15, FW37, FW43}, and also industry developments, such as Walmart's IBM Hyperledger Framework~\cite{FW24, FW3, FW23}, and SmartNoshWaste~\cite{FW6}. 

The frameworks proposed to reduce food waste are Hyperledger Fabric~\cite{FW3}, SmartNoshWaste~\cite{FW6}, Hyperledger Sawtooth~\cite{FW38}, or consortium blockchain~\cite{FW12}. Public Blockchain~\cite{FW14, FW15, FW36} and IBM Hyperledger Framework~\cite{FW4, FW23, FW24} were used to create solutions to ensure food safety. As food security remains a concern, public blockchain have been proposed to address it~\cite{FW43, FW7}. To improve the existing blockchain application in the food supply chain,  public blockchain~\cite{FW37} and PBFT~\cite{FW41} were considered.

Hyperledger Sawtooth is employed to reduce food waste~\cite{FW38}. Since all stakeholders can identify any participants along the supply chain, this approach increases the degree of trust among participants.

Walmart IBM Hyperledger Framework is used to ensure food safety in a practical implementation~\cite{FW4, FW23, FW24}. Proof of concepts (POCs), as a proprietary consensus scheme, ensures that blockchain provides a mechanism to trace and authenticate food as it goes from farm to store with speed and precision, mainly with a focus on traceability and authenticity.

\subsubsection{Supporting Technologies}

Supporting technologies such as using RFID tags \cite{FW15,FW43,FW37} and other IoT devices~\cite{FW37,FW45,FW36} to track the food product along the supply chain have been widely adopted in proposed blockchain systems that address food waste. 
For instance, QR codes~\cite{FW3,FW6,FW38}, RFID~\cite{FW3}, and cloud computing~\cite{FW6} are common supporting technologies in such works. 
QR codes and RFID provide a way to track the entire path the raw dairy travels to reduce dairy waste. 
Cloud computing can improve data traceability and accessibility to every stakeholder, including the consumer, which can reduce the waste generated by a product of unknown origin~\cite{FW6}.

IoT~\cite{FW36} and RFID~\cite{FW15} are used for traceability, finding food provenance and enhancing food safety. 
These IoT devices are used in farms, processing plants, plantation fields, and logistics companies to digitize the system.
Customers can verify the information for the products they want to purchase by scanning the RFID tags~\cite{FW15}. 

Similarly, IoT~\cite{FW45} devices and RFID~\cite{FW43} are used to strengthen food security by providing information regarding food location and the history of the product along the supply chain. 
Furthermore, the result of ML classification methods presents food-related information to help the stakeholders better manage their supply chain~\cite{FW43}. 

A multi-agent system architecture was proposed to improve food supply chain~\cite{FW11}.
The architecture has five layers: producer, processor, blockchain, retailer, and transport, to keep track of the product and ensure data integrity for an improved supply chain.

\subsubsection{ Challenges of Existing Blockchains}

Food provenance is one of the main concerns for consumers~\cite{FW_7, FW4}. Transparency and traceability, as indirect features of blockchain, can help to satisfy the demand for information provenance. Because the food supply chain is complex and extensive in terms of the number of products and processes, and it involves many stakeholders, blockchain's scalability remains a key challenge. Furthermore, data governance is another challenge because the private sector mostly manages food systems. For example, determining who is eligible to access data along the food supply chain remains debatable. 

Although blockchain and smart contracts can help users navigate the intricacies of regulation and compliance for the food supply chain systems, how they can be standardized and implemented across the numerous stakeholders along the supply chain remains an open issue.  

\subsubsection{Other Existing Surveys on Food Waste Management}

The surveys on blockchain applications in the food supply chain focus on different phases and challenges of the supply chain.
A gap of blockchain integrated with IoT in e-agriculture is bridged~\cite{FW1}.
In hospitality industry, impact of blockchain on food waste is evaluted~\cite{FW2}. 

Astarita et al.~\cite{FW5} reviewed the main challenges blockchain has in the transportation of food, the origin of food products, and trust in information reported in the supply chain. 
W{\"u}nsche et al.~\cite{FW9} evaluated the relationship between decreased food waste and enhanced working conditions throughout the supply chain.

Rejeb et al.~\cite{FW22} presented a review of the challenges, future work, and possible theoretical approaches and implementations focusing on food traceability, collaboration within the food supply chain, and food trading.

Kamilaris et al.~\cite{FW35} presented an overview of the impact of blockchain technology in agriculture and the food supply chain. 
Duan et al.~\cite{FW16} explored the possibilities of using blockchain to improve food traceability, information transparency, and recall efficiency. They also discussed the potential challenges blockchain systems might have in implementing these applications, such as raw data manipulation, stakeholder recruiting, and regulations.

Pakseresht et al.~\cite{FW19} analyzed the role of the circular economy and its applications in food safety. 
Saha et al.~\cite{FW10} reviewed 55 papers published from 2018 to 2022 on food waste and traceability in terms of the major companies that use blockchain in the food supply chain. Yogarajan et al.~\cite{FW21} presented a similar study on research trends, gaps, and future directions of applying blockchain in agri-industry from 27 papers. 

Different from the existing surveys, this survey focuses on a holistic application of blockchain in improving the food supply chain to reduce food waste and ensure food safety and security. 

\subsection{Water Management}
\label{sec:water-management}

Water scarcity is a pressing and recurring issue that countries worldwide face~\cite{water}. Due to climate change and extreme weather events, securing the availability and quality of the water resource is critical. However, water management, in general, suffers from issues such as information asymmetry~\cite{WT9}, staff shortage~\cite{WT10}, and data tampering~\cite{WT10}. To solve these issues, some research groups proposed using blockchain technology, which utilizes decentralization and immutability as well as other features for water management. The literature on blockchain for water management focuses on three objectives: improve water management, support water trading, and conserve water resources. 
Satilmisoglu et al.~\cite{surv_water} showed that blockchain can help reduce water waste, preserve water rights, and contribute to water sustainability.

\begin{figure}[htp!]
    \centering
\includegraphics[width=0.95\columnwidth]{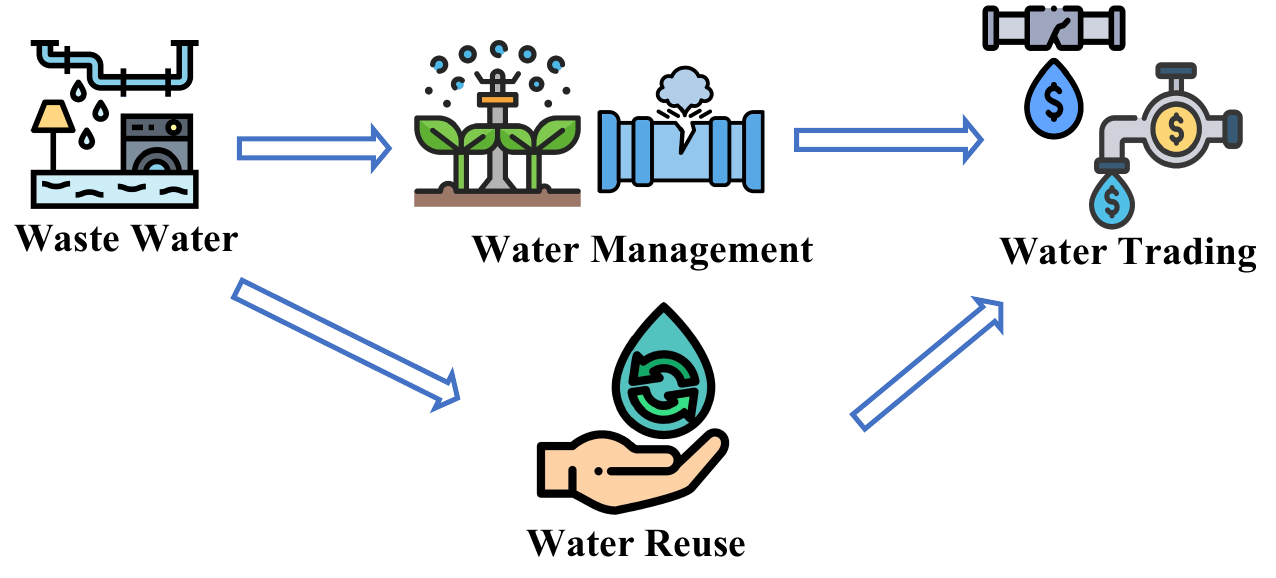}
    \caption{Objectives of blockchain applications on water management.}
    ~\label{fig:water}
\end{figure}

In this section, we overview the objectives pursued by existing blockchain applications on water management. We identify their targeted water sector, influencing stakeholders, utilizing blockchain features, employing blockchain frameworks, the consensus algorithm, supporting technologies, and challenges. Figure \ref{fig:water} shows the objectives of blockchain applications on water management. Table \ref{tab:water} shows the classification of the surveyed blockchain applications on water management according to their objectives.

\begin{table*}[t!]
  \caption{Blockchain applications on water management.}
   \renewcommand{\arraystretch}{1}
    \begin{tabular}{|>{\centering\arraybackslash}p{1.3cm}|>{\centering\arraybackslash}p{1.6cm}|@{}>{\centering\arraybackslash}p{2.7cm}@{}|@{}>{\centering\arraybackslash}p{0.4cm}@{}|@{}>{\centering\arraybackslash}p{0.4cm}@{}|@{}>{\centering\arraybackslash}p{0.4cm}@{}|@{}>{\centering\arraybackslash}p{0.4cm}@{}|@{}>{\centering\arraybackslash}p{0.4cm}@{}|@{}>
    {\centering\arraybackslash}p{0.4cm}@{}|@{}>{\centering\arraybackslash}p{0.4cm}@{}|@{}>
    {\centering\arraybackslash}p{0.4cm}@{}|@{}>{\centering\arraybackslash}p{2.3cm}|>{\centering\arraybackslash}p{2.3cm}|>{\centering\arraybackslash}p{2cm}|}\hline

    \multicolumn{1}{|c|}{\multirow{2}{*}{\bf Objective}}&
    \multicolumn{1}{c|}{\multirow{2}{*}{\makecell{\bf Sector}}}&
    \multicolumn{1}{c|}{\multirow{2}{*}{\makecell{\bf Influencing\\ \bf Stakeholders}}}&
    \multicolumn{8}{c|}{\bf Blockchain Features}&
    \multirow{2}{*}{\makecell{\bf Framework\\ \bf (Consensus)}}&
    \multirow{2}{*}{\makecell{\bf Supporting\\ \bf Technologies}}&
    \multirow{2}{*}{\makecell{\bf Challenges}}\\\cline{4-11}

    & & & 
    \textbf{\scriptsize SC}&
    \textbf{\scriptsize DS}&
    \textbf{\scriptsize IN}&
    \textbf{\scriptsize PY}&
    \textbf{\scriptsize TC}&
    \textbf{\scriptsize TK}&
    \textbf{\scriptsize TP}&
    \textbf{\scriptsize TR}& & & \\
    \hline
    {\multirow{16}{*}{\makecell{Improve \\water\\ manage-\\ment}}}
    & \multirow{5}{*}{\makecell{Wastewater \\~\cite{WT57,WT10}\\ \cite{WT6,WT11,WT13,WT19,WT40,WT53}}}
    & {Service providers, government, environmental monitoring agencies, NGOs, water suppliers and distribution companies}
    & \multirow{5}{*}{{\scriptsize \checkmark}}& \multirow{5}{*}{{\scriptsize \checkmark}}& \multirow{5}{*}{{\scriptsize \checkmark}}& {}& \multirow{5}{*}{{\scriptsize \checkmark}}& \multirow{5}{*}{{\scriptsize \checkmark}}& \multirow{5}{*}{{\scriptsize \checkmark}}& \multirow{5}{*}{{\scriptsize \checkmark}}
    & \multirow{5}{*}{\makecell{Hyperledger \\Fabric \\(Kafka)~\cite{WT6}, \\double \\blockchain~\cite{WT40},\\(DPoS)~\cite{WT13}\\ (PoSL~\cite{WT57}), \\(PoML~\cite{WT57})}}
    & \multirow{5}{*}{\makecell{ML~\cite{WT6},\\~\cite{WT40},\\ IoT~\cite{WT6,WT57}, \\ \cite{WT40},\\IIoT~\cite{WT10}}}
    & \multirow{10}{*}{\makecell{Scalability\\~\cite{WT6}}}\\
     \cline{2-13}
    
    {}
    & \multirow{5}{*}{\makecell{Harvested \\water \\~\cite{WT9, WT27}\\~\cite{  WT12,WT16, WT46, WT58, WT59}}}
    & {Governments, industries, universities, NGOs, and groundwater sustainability agencies}
    & \multirow{4}{*}{\scriptsize \checkmark}& \multirow{4}{*}{\scriptsize \checkmark}& {}& {}& \multirow{4}{*}{\scriptsize \checkmark}& {}& \multirow{4}{*}{\scriptsize \checkmark}& {}
    & \multirow{5}{*}{\makecell{(PoAh)~\cite{WT27},\\ Ethereum~\cite{WT46}, \\Hyperledger \\Fabric\\~\cite{WT58}}}
    & \multirow{5}{*}{\makecell{ IoT~\cite{WT16, WT27},\\\cite{WT59},\\ IoUT~\cite{WT27},\\ IoAT~\cite{WT46},\\IPFS~\cite{WT58}}}
    & {}\\
    \cline{2-13}

    {}
    & \multirow{9}{*}{\makecell{ Irrigation\\~\cite{WT1, WT7, WT8, WT15, WT17,WT22, WT31, WT42,WT44, WT48, WT50,WT55,WT60}}}
    & \makecell{Farms, consumers, \\agriculture \\agencies, \\ system owners\\}
    & \multirow{5}{*}{\scriptsize \checkmark}& \multirow{5}{*}{\scriptsize \checkmark}& {}& {}& \multirow{5}{*}{\scriptsize \checkmark}& \multirow{5}{*}{\scriptsize \checkmark}& \multirow{5}{*}{\scriptsize \checkmark}& \multirow{5}{*}{\scriptsize \checkmark}
    & {Ethereum~\cite{WT1, WT7, WT17}, Hyperledger Fabric~\cite{WT44}, (PoW)~\cite{WT1}, Alliance chain (PBFT)~\cite{WT22}, PoST~\cite{WT42}}
    & {IoT~\cite{WT15,WT17, WT48}, IoUT~\cite{WT48}, IoWT~\cite{WT60}, sensors~\cite{WT1, WT48}, RFID~\cite{WT7}, LPWAN~\cite{WT15}, Fuzzy logic~\cite{WT8}}&
    \\
    \cline{2-13}
  
    {}
    & \multirow{4}{*}{\makecell{Municipal\\~\cite{WT14, WT30, WT33, WT36, WT39, WT43, WT56}}}
    & \makecell{Agencies, \\ water supply\\ plants}
    & \multirow{5}{*}{\scriptsize \checkmark}& \multirow{5}{*}{\scriptsize \checkmark}& {}& {}& \multirow{5}{*}{\scriptsize \checkmark}& \multirow{5}{*}{\scriptsize \checkmark}& \multirow{5}{*}{\scriptsize \checkmark}& {}
    & {Private Blockchain~\cite{WT39}, Hyperledger Fabric~\cite{WT36} }
    & {Smart meter~\cite{WT43,WT33, WT36, WT39}, ML~\cite{WT14}, IoT~\cite{WT36,WT33, WT14, WT39}, Bloom filter algorithm~\cite{WT39}}
    & \\ 
    \hline

    \multirow{5}{*}{\makecell{Support\\ Water\\ Trading}} 
    & \multirow{5}{*}{\makecell{Agriculture\\~\cite{WT1} \\Municipal\\~\cite{WT3}}}
    & {Government, water conservancy enterprises, third-party maintenance, households, and irrigators}
    & \multirow{5}{*}{\scriptsize \checkmark}& {}& {}& \multirow{5}{*}{\scriptsize \checkmark}& \multirow{5}{*}{}& \multirow{5}{*}{\scriptsize \checkmark}& \multirow{5}{*}{\scriptsize \checkmark}& {}
    & {Ethereum (PoW)~\cite{WT1}, Ethereum (PoA)~\cite{WT3}}
    & \multirow{5}{*}{\makecell{WSN~\cite{WT3},\\ sensors~\cite{WT1},\\ \cite{WT3}}}
    & {Computational and storage overheads, high latency~\cite{WT3}}\\
    \hline
    
    \multirow{5}{*}{\makecell{Conserve\\ Water\\ Resources}}
    & \multirow{3}{*}{\makecell{Municipal\\~\cite{WT5,WT52}}}
    & {Government, institutions}
    & \multirow{5}{*}{\scriptsize \checkmark}& {}& {}& \multirow{5}{*}{\scriptsize \checkmark}& {}& {}& \multirow{5}{*}{\scriptsize \checkmark}& {}
    & {Hyperledger Fabric~\cite{WT52}}
    & IoT~\cite{WT5, WT52}, cloud computing~\cite{WT5}
    & {Energy consumption, lack of regulation~\cite{WT20, WT23}}\\
    \cline{2-13}
    {}
    & \multirow{4}{*}{\makecell{Harvested \\water~\cite{WT47},\\~\cite{WT23,WT26,WT20}}}
    & \makecell{Government \\agencies, \\industries, \\households}
    & {}& \multirow{2}{*}{\scriptsize \checkmark}& {}& {}& \multirow{2}{*}{\scriptsize \checkmark}& {}& \multirow{2}{*}{\scriptsize \checkmark}& {}
    & {Ethereum~\cite{WT26}, (PoW)~\cite{WT20}, Alliance chain~\cite{WT23}, Corda~\cite{WT47}}
    & User interface~\cite{WT47}
    & {}\\
    \hline

    \end{tabular}%
  \label{tab:water}%
\end{table*}%

\subsubsection{Objectives of the use of Blockchain in Water Management}

\paragraph{Improve Water Management} 

The majority of the literature focuses on designing blockchain solutions to improve water management that encompasses different water sectors, including wastewater, harvested water (rain, river, groundwater), irrigation water, and municipal water. 

Wastewater is the water discharged from industry and households~\cite{WT6}. Direct wastewater disposal potentially pollutes the water and damages aquatic life~\cite{WT13}. Thus, it is important to treat wastewater. Iyer et al.~\cite{WT6} proposed a blockchain-based wastewater treatment management system. 
These works propose conceptual architectures of blockchain-based wastewater treatment management systems and discuss requirements for such systems. 

Efficient water resource management for more effective water utilization using blockchain is also a matter of interest. Xia et al.~\cite{WT19} explored a blockchain framework for intelligent water management to ensure secure and verifiable water abstraction permit and license information. They proposed a blockchain-based water resource information management system to mitigate the information asymmetry between the government and water users. This blockchain-based system can monitor water use and trace the water quality. Mahmoud et al.~\cite{WT40} proposed WDSchain, a blockchain-IoT toolkit that evaluates water distribution systems while preserving the data security of the input information. They also proposed a combination of ML and the consensus algorithms PoSL and PoML to validate the data input by the IoT devices~\cite{WT57}. Alharbi et al.~\cite{WT53} proposed a blockchain-based water quality monitoring system to monitor wastewater from the industries in Saudi Arabia to ensure compliance.

Harvested water management centers on collected water from rain, rivers, and ground. Water management using IoT or Industrial Internet of Things (IIoT) systems is also called intelligent water management (IWM)~\cite{WT12}. The challenges that IWM faces are water loss due to pipe bursts, depletion of resources, and water pollution. To solve IWM problems such as high cost, low efficiency, and insecure data storage due to centralization,
Chohan et al.~\cite{WT9} proposed IBM’s blockchain-based water management system to reduce information asymmetry for groundwater usage. An increased interest is in exploring how blockchain improves groundwater systems from a sustainable~\cite{WT12}, efficient~\cite{WT27}, or trading perspective~\cite{WT59}.
Tajudin et al.~\cite{WT16} proposed a blockchain system to secure data collected from IoT devices. They measure water quality to prevent pollution generated by waste dumping on rivers, in Malaysia. Ali et al.~\cite{WT58} proposed a blockchain-based recording system for river stream flow data with IPFS off-chain storage to assist in decision-making in irrigation and flood mitigation systems. The approach also aims to increase interoperability among these stakeholders in Pakistan.

Water management for irrigation focuses on the governance of water usage for agriculture purposes~\cite{WT7,WT8,WT15,WT17,WT22,WT31,WT42,WT48,WT50,WT55,WT60}. IoT, IIoT, and Industrial Internet of Water Things (IIoWT) devices are commonly employed to monitor water usage for agriculture.
The water crisis is a common problem faced by the irrigation community. To solve this ongoing problem, a few blockchain-based water control systems are proposed to manage water for irrigation communities~\cite{WT7, WT31, WT42, WT50}, to mitigate water stress in agriculture~\cite{WT15, WT17}, and to provide real-time information for water right trading~\cite{WT22}. Blockchain ensures data security in monitoring water consumption for medium-scale gardens and fields~\cite{WT8}. Rewards or penalty proposals using blockchain have also been considered for leveraging incentives to conserve water~\cite{WT31, WT42}. Another application is the use of blockchain to monitor data on photovoltaic panels and water pumps. SolarCoin is an example of using tokens to create new energy and water trading opportunities~\cite{WT42}. Pincheira et al.~\cite{WT31} proposed a blockchain-based water management system using IoT devices to test the feasibility of integrating off-the-shelf hardware and providing sensed data to incentivize sustainable behavior in irrigation water. In such work, the information from the IoT devices is recorded in smart contracts to preserve transparency and accuracy.

Municipal water management focuses on water use for residential, industrial, and smart cities~\cite{WT14, WT30, WT33, WT36, WT39, WT43}. IoT and smart meters are commonly employed to monitor water usage. 
Monitoring household water use is a key measure for improving water usage within municipalities. With a similar objective, Thakur et al.~\cite{WT14} proposed a blockchain-based system to incentivize water saving by households in a municipality. In this work, rewards are distributed to households likely to save water using smart contracts. The model uses machine-learning-prediction and historical water usage. 
Bracciali et al.~\cite{WT33} proposed a blockchain-based system to vote on water policies. There is an increasing focus on using blockchain to monitor real-time data on residential water consumption~\cite{WT39, WT36} and ensuring that information passed through smart meters is not altered~\cite{WT43}.

\paragraph{Support Water Trading}
The current water trading market is challenged by the lack of transparency between stakeholders, administrative complexities, and complicated financial settlement processes. 
The lack of transparency among stakeholders and complicated financial settlement processes challenges the current water trading market. Pee et al.~\cite{WT1} proposed a blockchain-based water trading system that ensured transparency between sellers and buyers and enabled immediate payments using smart contracts.
Blockchain was employed for water trading and water rights exchanges~\cite{WT3}.

\paragraph{Conserve Water Resource}

Conserving~\cite{WT5, WT52} and protecting water against pollution~\cite{WT20, WT26, WT23, WT47} have motivated the adoption of blockchain in water management. Dogo et al.~\cite{WT5} reviewed how blockchain can improve the use of African water resources. 
Shi et al.~\cite{WT52} proposed an IoT blockchain system that was implemented in 39 Chinese schools to protect the quality of drinking water by ensuring that the recorded status of the drinking water is not tampered with. Wu et al.~\cite{WT20} analyzed the application, challenges, and limitations of utilizing blockchain technology for water resource protection.  Zhang et al.~\cite{WT23} proposed a conceptual water information-sharing platform that allows participants to trade water rights. Blockchain-based water rights trading was proposed to ensure data security. 
Niya et al.~\cite{WT26} proposed a design of a blockchain-based pollution monitoring system to preserve data from IoT devices that detect PH, turbidity for water, carbon dioxide, and carbon monoxide in the environment.
Crawford et al.~\cite{WT47} proposed a permissioned blockchain to allow oil and gas companies to create, disseminate, and trace immutable records with state agencies to protect aquifers in California.

\subsubsection{Influencing Stakeholders}

Blockchain-based water management systems involve a wide range of stakeholders, including government agencies at different levels that set water policies and check for compliance, manufacturers of monitoring devices, water suppliers and distributors, blockchain solution providers, non-governmental organizations, industrial plants that generate and treat wastewater, agriculture farms, irrigation communities, and individual users. 

\subsubsection{Blockchain Features Sought in Water Management}

Blockchain-based water management systems are interested in enhancing transparency to reduce water information asymmetry between stakeholders, enabling better water use cooperation. The access to consistent data and secure payment enabled by smart contracts also allows entities to trade water abstraction rights. The water management system also uses blockchain tracking to find locations with inadequate water quality and traceability features to trace water quality.

\subsubsection{Blockchain Frameworks used in Water Management}

Blockchain-based water management systems are implemented on Hyperledger Fabric~\cite{WT58, WT53, WT6, WT36, WT44, WT52}, Alliance chain~\cite{WT22, WT20, WT23}, Ethereum~\cite{WT1, WT3, WT7, WT17, WT46}, Ethereum Light Client (ELC)~\cite{WT26}, and R3 Corda~\cite{WT47}. Hyperledger Fabric was used to check the feasibility of real-time data storage~\cite{WT36}. A system using a static blockchain for one-time intervals and two dynamic blockchains for time series was proposed considering different security levels based on the consensus (PoW, PoT, or PoV) used by the network~\cite{WT40}. A private blockchain uses k-means++ to group users into clusters where each cluster has a private blockchain to record the data of members~\cite{WT39}.

\subsubsection{Supporting Technologies} 
IoT devices are commonly deployed to collect data for
wastewater treatment management~\cite{WT6, WT10, WT13, WT11, WT19, WT40, WT57, WT53}, harvested water~\cite{WT9,WT12, WT16, WT27, WT46, WT59, WT58}, agricultural water management~\cite{WT1, WT7, WT8, WT15, WT17, WT22, WT31, WT42, WT44,  WT48, WT50, WT55, WT60}, and municipalities~\cite{WT36,WT33, WT14, WT39, WT30, WT43, WT56}. IIoT, developed for industrial applications, is deployed to wastewater treatment management where the risk impact factor is high~\cite{WT10}. Internet of Underwater Things (IoUT)~\cite{WT27} and Internet of Agricultural Things (IoAT)~\cite{WT46} are used as specific hardware to monitor water. The IIoWT was considered for data standardization, interoperability, and data security among different water institutions to be met~\cite{WT60}.
IoT~\cite{WT6, WT57, WT40, WT16, WT27, WT5, WT15, WT17, WT48, WT36, WT33, WT14, WT39, WT52} and IIoT~\cite{WT10} devices can monitor water use and directly interact with blockchain through Low Power Wide Area Networks (LPWAN)~\cite{WT15}.
ML techniques predict water use~\cite{WT14} and detect anomalies in data~\cite{WT6}. For agricultural water management, RFID readers are adopted to grant users water access according to the rules set by the community, also referred to as irrigation receipt~\cite{WT7}.

\subsubsection{Challenges of Existing Blockchains} 

The identified challenges of blockchain for water management from a number of the reviewed literature are policy and regulation-based~\cite {WT19,WT20}. Considering that water rights and access to water often cross national borders, the implementation of blockchain for water management needs to consider international laws and national policies, as well as geopolitical cases, to allow collaboration and cooperation amongst stakeholders across borders. 

Technical challenges that blockchain-based systems face are the energy consumption of the blockchain application~\cite{WT20,WT23} and
the scalability of the system~\cite{WT3,WT6}, especially under a massive number of IoT devices and sensors. 
Transparency of information could facilitate informed decision-making and encourage collaboration. The cost and complexity of monitoring the vast water systems require significant resources and collaboration. A feasibility study could be an important first step towards filling this gap.  

\subsection{Circular Economy and Blockchain}
\label{sec:circular-economy}

\begin{table*}[t!]
    \centering
    \caption{Blockchain applications for management of circular economy.}
    \renewcommand{\arraystretch}{1}

    \begin{tabular}{|>{\centering\arraybackslash}p{1.6cm}|>{\centering\arraybackslash}p{3.5cm}|>{\centering\arraybackslash}p{0.3cm}|>{\centering\arraybackslash}p{0.3cm}|>{\centering\arraybackslash}p{0.3cm}|>{\centering\arraybackslash}p{0.3cm}|>{\centering\arraybackslash}p{0.3cm}|>{\centering\arraybackslash}p{2.5cm}|>{\centering\arraybackslash}p{1.8cm}|>{\centering\arraybackslash}p{1.9cm}|}
    \hline

    \multirow{2}{*}{\textbf{Objective}}&
    \multirow{2}{*}{\makecell{\bf Influencing\\ \bf stakeholders}}&
    \multicolumn{5}{c|}{\textbf{Blockchain Features}}&
    \multirow{2}{*}{\makecell{\bf Blockchain \\ \bf Framework}}&
    \multirow{2}{*}{\makecell{\bf Supporting\\ \bf Technologies}}&
    \multirow{2}{*}{\makecell{\bf Challenges}}\\
    \cline{3-7}

   & &\textbf{\scriptsize SC} & \textbf{\scriptsize DG}& \textbf{\scriptsize TC}& \textbf{\scriptsize TK}& \textbf{\scriptsize TP}& & &\\ \hline
    
    \multirow{10}{*}{\makecell{Circulate \\supply chain \\product\\~\cite{CE4,CE6,CE7,CE8,CE18,CE28,CE13}}}&
    General supply chain: government, business, regulator, producer, builder, trader, etc. \cite{CE4,CE6,CE8,CE28,CE13}&
    \multirow{4}{*}{\scriptsize{\checkmark}} & 
    \multirow{4}{*}{\scriptsize{\checkmark}} & 
    \multirow{4}{*}{\scriptsize{\checkmark}}& 
    {}& 
    \multirow{4}{*}{\scriptsize{\checkmark}}& 
    {Hyperledger Fabric~\cite{CE28}}& {N/A} & \multirow{10}{*}{\makecell{DApp for\\ stakeholder\\ interaction\\ with \\blockchain \\\cite{CE6}, \\scalability \\\cite{CE8}}} \\
   \cline{2-9} 
    &Energy sector: power plant, energy provider, distribution operator, prosumer, consumer \cite{CE7}&
    \multirow{4}{*}{\scriptsize{\checkmark}} & 
    \multirow{4}{*}{\scriptsize{\checkmark}}&
    \multirow{4}{*}{\scriptsize{\checkmark}}& 
    {}& 
    \multirow{4}{*}{\scriptsize{\checkmark}}& 
    {N/A}&
    {N/A} &\\
    \cline{2-9}
    &
    Building sector: producer, regulator, builder \cite{CE18} &
    \multirow{2}{*}{\scriptsize{\checkmark}} & 
    \multirow{2}{*}{\scriptsize{\checkmark}}&
    \multirow{2}{*}{\scriptsize{\checkmark}}& 
    {}& 
    \multirow{2}{*}{\scriptsize{\checkmark}}& 
    {Hyperledger Fabric}& {N/A} &\\
    \hline

    {Recycle e-waste~\cite{CE19,CE26}}& {Chip company, authenticator, computer company, customer, authorized recycler} & 
    \multirow{4}{*}{\scriptsize{\checkmark}} & 
    {}&
    \multirow{4}{*}{\scriptsize{\checkmark}}&
    \multirow{4}{*}{\scriptsize{\checkmark}}&
    \multirow{4}{*}{\scriptsize{\checkmark}}&
    {Algorand blockchain (implemented)~\cite{CE19}, Hyperledger Fabric~\cite{CE26}}& {IoT~\cite{CE26}} & {Scalability \cite{CE19} security and privacy policy \cite{CE26}}\\
    \hline

    {Reuse plastic~\cite{CE20,CE30,plastics_PR7}}& {Plastic manufacturers, government entities, retailers, suppliers, waste collectors, and recyclers} &
    \multirow{4}{*}{\scriptsize{\checkmark}} & \multirow{4}{*}{\scriptsize{\checkmark}}& \multirow{4}{*}{\scriptsize{\checkmark}}& \multirow{4}{*}{\scriptsize{\checkmark}}& \multirow{4}{*}{\scriptsize{\checkmark}}& {Hyperledger Fabric (implemented)~\cite{CE20}}& {ML~\cite{CE30}, sensors~\cite{plastics_PR7}} & {Interoperability \cite{CE20}}\\
    \hline
      
\end{tabular}%
\label{tab:ce}%
\end{table*}%

\begin{table*}[t!]
    \centering
    \caption{Survey papers on blockchain for circular economy.}
    \renewcommand{\arraystretch}{1}

    \begin{tabular}{|>{\raggedright\arraybackslash}m{6cm}|>{\centering\arraybackslash}p{0.3cm}|>{\centering\arraybackslash}p{0.3cm}|>{\centering\arraybackslash}p{0.3cm}|>{\centering\arraybackslash}p{0.3cm}|>{\centering\arraybackslash}p{0.3cm}|>{\centering\arraybackslash}m{2.3cm}|>{\centering\arraybackslash}m{3.9cm}|}
    \hline

    \multicolumn{1}{|c|}{\multirow{2}{*}{\textbf{Objective}}}&
    \multicolumn{5}{c|}{\textbf{Blockchain Features}}&
    \multicolumn{1}{c|}{\multirow{2}{*}{\makecell{\bf Supporting \\ \bf Technologies}}}&
    \multicolumn{1}{c|}{\multirow{2}{*}{\textbf{Challenges}}}\\
    \cline{2-6}

    & \textbf{\scriptsize{SC}}& \textbf{\scriptsize DG} &\textbf{\scriptsize TC}& \textbf{\scriptsize TK}& \textbf{\scriptsize TP}& & \\ \hline
    
    {Enhance sustainability and social responsibility~\cite{CE1}}&{\scriptsize{\checkmark}} &{\scriptsize{\checkmark}} & & & &{N/A}& {N/A}\\
    \hline

    {Create value chain, track material, access data, and build trust~\cite{CE2,CE5}}& {\scriptsize{\checkmark}}&{\scriptsize{\checkmark}}& {\scriptsize{\checkmark}}& {\scriptsize{\checkmark}}& {\scriptsize{\checkmark}}& {N/A}& {N/A}\\
    \hline

    {Use blockchain as an enabler for CE~\cite{CE3,CE10,CE11,CE14,CE21,CE22,CE24,CE25,CE27,CE32,baralla2023waste,jiang2023blockchain}}&{\scriptsize{\checkmark}} & & {\scriptsize{\checkmark}}& &{\scriptsize{\checkmark}} & {IoT~\cite{CE10,CE21,CE22,CE24}, RFID~\cite{CE22}}& {High energy cost~\cite{CE21,CE25}, scalability~\cite{CE25}}\\
    \hline

    {Perform gap analysis in using blockchain in industrial sectors~\cite{CE15,CE16}}&{\scriptsize{\checkmark}} & {\scriptsize{\checkmark}}& {\scriptsize{\checkmark}}& & {\scriptsize{\checkmark}}& {N/A} & {Scalability, human-related security loopholes, high energy cost~\cite{CE16}}\\
    \hline

\end{tabular}%
\label{tab:ce_survey}%
\end{table*}%

\begin{figure}[htp!]
    \centering
\includegraphics[width=0.95\columnwidth]{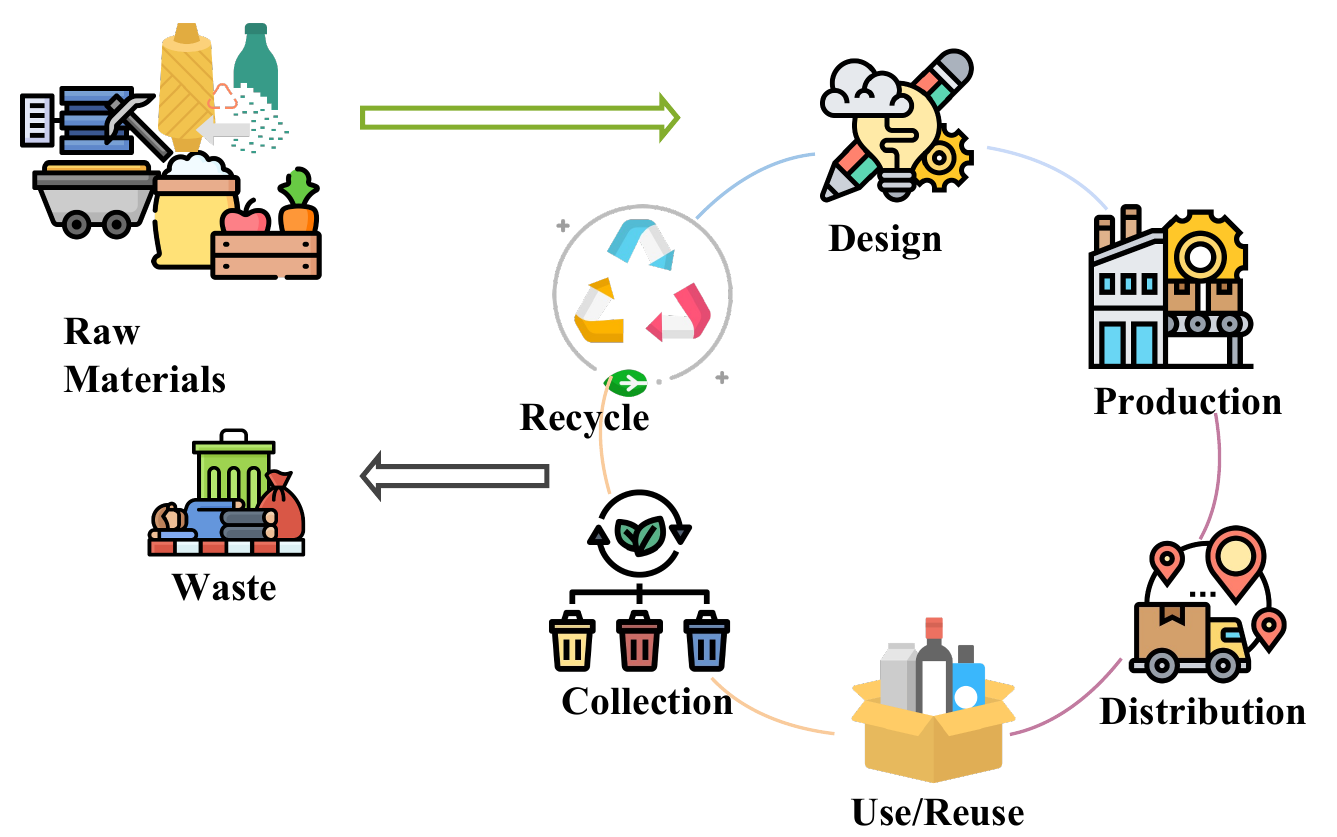}
    \caption{Representation of stages involved in circular economy.}
    ~\label{fig:ce}
\end{figure}

A circular economy (CE) is a mechanism that reuses, renews, and regenerates materials, products, or services to keep them as or make them more sustainable or environmentally friendly \cite{CE2}. CE is also a new business model applied to the design of products to make them reusable, with durable materials that can fulfill those purposes, and methods for extending the useful lifetime of products or materials. The reuse, renew, and regenerate approaches followed in CE have great potential to reduce waste for sustainable development \cite{CE3}. Figure \ref{fig:ce} shows an overview of the circular economy concept. CE is emerging as a sustainable alternative to the traditional linear economy system, where consumers buy a product, use it, and dispose of it \cite{CE1}.  
However, the implementation of a circular economy faces many economic and management challenges, such as tokenization of the resources used along the supply chain, mechanisms for tracking, and setting the value proposition of recycled materials and resources, among others \cite{CE1,CE3,CE5}. Therefore, there is an increasing interest in exploring blockchain as a solution to address some of these challenges. 

We review the existing works that use blockchain solutions for the circular economy, including case studies in various industries, and highlight future research needs. 
Table \ref{tab:ce} summarizes existing work on using blockchain for circular economy. Because the circular economy is still in its infancy, the majority of the reported work focuses on evaluating the potential overlaps of circular economy and blockchain while very little work has used implementations of blockchain. Table \ref{tab:ce_survey} summarizes existing surveys and case studies that evaluate the potential blockchain applications for different industries, overview the concepts and general gaps in a circular economy, and outline how blockchain features can be adopted to address them. 

\subsubsection{Objectives of using Blockchain in  Circular Economy}

\paragraph{Circulate supply chain products}

Most of the works on blockchain applications for the management of circular economy aim to address the question of how to circulate the products for reuse along the supply chain \cite{CE4,CE6,CE7,CE8,CE18,CE28}. Learning from the COVID-19 pandemic experience, Nandi et al. \cite{CE4} proposed to redesign a supply chain and use blockchain to achieve localization, agility, and digitization (LAD) where blockchain is used to enable digitization of supply chain records and to provide reliable data that tracks waste and inventory for localization and traceability to improve responsiveness (agility) for circular economy. 
Alexandris et al. \cite{CE6} presented a conceptual blockchain-based asset recording, accessing, and sharing mechanism that allows auditors and governmental regulators to monitor the state of the asset and its transfer between all parties with integrity. 
Kouhizadeh et al. \cite{CE8} presented the first conceptual analysis of using blockchain for product deletion in a circular economy. 
Triple Retry is also a blockchain-based CE framework that employs Hyperledger Fabric \cite{CE28}. In this framework, Crash Fault Tolerant (CFT), and BFT can be chosen based on the firm's throughput and speed requirements. In e-commerce, IoT and blockchain have crucial influences on the virtual supply chain. Aiming to achieve a virtual closed-loop supply chain (VCLSC), Prajapati et al. \cite{CE13} used a single mixed-integer non-linear programming model to maximize the total expected revenue from VCLSC. In this model, the costs of physically handling the materials and virtually running blockchain and IoT in VCLSC were both taken into account. The result can be used to guide business organizations and government administrators to create better policies in VCLSC.
Yildizbasi et al. \cite{CE7} integrated blockchain with renewable energy grid management to ensure the efficient distribution of energy, reduce illegal energy use, and allow small-scale renewable energy producers to supply to the energy grid. 
Shojaei et al. \cite{CE18} investigated the possibility of using blockchain for realizing CE in the building and construction sector, which is tested on Hyperledger Fabric. The result from a synthetic case study shows that this application can be used to track building materials, and that makes proactive planning of recycling and reusing materials possible.

\paragraph{Recycle electronic waste}

Recycling electronic waste or e-waste is another objective of blockchain implementation for circular economy \cite{CE19,CE26}. Taking into account different parties in the supply chain, CircleChain, a role-based token management scheme, is proposed to authenticate, synthesize, circulate, and reuse second-life components in a trust-less environment \cite{CE19}. It was implemented on Algorand blockchain and its scalability was tested using smart contracts and Algorand Standard Assets (ASA). 
Hatzivasilis et al. \cite{CE26} proposed a blockchain-based CE framework where IoT logs the state of assets on the blockchain and ensures it is immutable. The approach also includes the proposal of federated learning to distribute the computationally heavy CE business tasks to conserve resources. 

\paragraph{Reuse plastics}

Plastic reuse is another primary application of blockchain in CE \cite{CE20,CE30,plastics_PR7}. 
PlasticCoin is an ERC20-compliant token implemented on Hyperledger Fabric \cite{CE20}. ERC20 is a standard on the Ethereum blockchain, which defines how tokens can be used by smart contracts. Participants in CE are rewarded with tokens when they collect, recycle, and reuse plastic waste. The tokens can be traded for goods. Two conceptual blockchain-based CE frameworks for plastic reuse were proposed \cite{CE30, plastics_PR7}. These frameworks predict the global plastic generation using ML techniques \cite{CE30} and segregate commingled plastic using sensor data \cite{plastics_PR7}. The predicted amounts can be considered for re-use planning and management.

\subsubsection{Influencing stakeholders}
In general, for CE blockchain solutions for circulating supply chain products, the influencing stakeholders include all parties along the supply chain, such as government, business, regulator, producer, trader, etc. For more specific applications in the different case studies, the influencing stakeholders are more clearly identified, e.g. chip company, authenticator, computer company, customer, authorized recycler are identified for recycling e-waste and plastic manufacturers, government entities, retailers, suppliers, waste collectors, and recyclers are identified for plastic reuse.

\subsubsection{Blockchain Features Sought in Circular Economy}

Smart contracts, traceability, and transparency are blockchain features sought in CE for the tracing of products, recycling, and reusing second-life materials. Given the complexity of the circular supply chain, data governance emerges as a crucial aspect, as identified in studies on the circulation of supply chain products and the management of plastic reuse to control data access. Tokenization is often reported as a feature in implemented systems, such as CircleChain and PlasticCoin, which target the recycling of e-waste and plastics, respectively.

\subsubsection{Blockchain  Frameworks used in Circular Economy}
Despite the limited work on CE, Hyperledger fabric is the most popular blockchain framework reported for CE applications~\cite{CE18,CE20,CE26,CE28}.
Algorand, a PoS fast-consensus blockchain platform, has been reportedly adopted for the implementation of CircleChain~\cite{CE19}. 

\subsubsection{Supporting Technologies}
The circular economy includes a large number of processes and stages in the life cycle of a product or material. Therefore, it requires a large number and variety of sensors and data input devices, processing, and recording. Some of those reported are IoT devices \cite{CE26}, sensors \cite{plastics_PR7}, and ML techniques \cite{CE30} that are used to keep track of second-life products and provide decision support based on the collected data. 

\subsubsection{Challenges in Blockchain Solutions for Circular Economy}

Scalability remains one commonly identified challenge for blockchain solutions for CE applications \cite{CE8,CE19}. However, easy-to-use interfaces such as DApp are needed to allow stakeholders to interact with the blockchain \cite{CE6}. Additionally, establishing security and privacy policy \cite{CE26} must be considered together with leveraging support for interoperability of different systems that rely on the required system features \cite{CE20}. 

\subsubsection{Existing Surveys on Blockchain-enabled CE}

There is a large number of surveys on CE in the literature. However, the main goals are not on blockchain but rather on other aspects of CE, such as sustainability advantages, social responsibility, value chain, and others.
Table \ref{tab:ce_survey} summarizes the focus points of existing surveys. They are categorized according to their objectives as follows: 
\begin{itemize}
\item Enhance sustainability and social responsibility \cite{CE1}: reduce transaction costs, enhance performance and communication along the supply chain, ensure human rights protection, enhance healthcare patient confidentiality and welfare, and reduce carbon footprint.
\item Create value chain, track material, access data, and build trust \cite{CE2,CE5}.
\item Use blockchain as an enabler for CE \cite{CE3,CE10,CE11,CE14,CE21,CE22,CE24,CE25,CE27,CE32}. 
\item Perform gap analysis in using blockchain in industrial sectors \cite{CE15, CE16}.  
\end{itemize}

 

These surveys discuss drivers and barriers to the adoption of blockchain for CE in different economic sectors. 
Rejeb et al. \cite{CE11} identified 19 enablers of employing blockchain for CE. 
The results showed that transparency, security, smart contracts, traceability, and enhanced collaboration are the critical causal enablers while immutability, decentralization, privacy, automation, information sharing, and enhanced regulation are the effect enablers. 
Pakseresht et al. \cite{CE22} discussed the role of blockchain in the transition towards a circular food system in the agri-food sector. Yontar et al. \cite{CE27} identified and ranked, by influence, the critical success factors.
Lack of knowledge and management support, reluctance to adopt new technology, and lack of technological progress are the top three barriers identified~\cite{CE16}.
 
\subsubsection{Challenges}

Although the circular economy has great potential for managing the sustainable use of resources with minimum impact on the environment, this new concept faces numerous challenges for its actual implementation. It requires an overhaul of the traditional product design process to consider materials that can be reused, repaired, and repurposed. Financial incentives for businesses to adopt the sustainable circular product design are needed as well as new business models that encourage collaboration and cooperation among various stakeholders. 

Blockchain has the potential to provide solutions for tokenization, transparency, traceability, data immutability, and automation of the process. However, how to balance security and scalability to support a large-scale circular economy system requires future study. 
More critically, most of the blockchain proposals for a circular economy are theoretical. Therefore, the cost and ease of implementation and whether business incentives can make the public embrace a circular economy are not known. While the concept seems to pack great not only sustainable but also economic benefits, a suitable implementation framework is required.
The high energy cost for the implementation of blockchain is also of concern for achieving sustainable development with energy-intensive computational infrastructure. This topic requires further exploration.

\section{Discussion and Future Challenges}
\label{sec:discussion}
\begin{table*}[t!]
    \centering
    \caption{Implementations of blockchain with available source code.}
    \renewcommand{\arraystretch}{1}

    \begin{tabular}{|>{\raggedright\arraybackslash}p{2.6cm}|>{\raggedright\arraybackslash}p{2.6cm}|@{}>{\centering\arraybackslash}p{1.75cm}@{}|@{}>{\centering\arraybackslash}p{1.75cm}@{}|>{\centering\arraybackslash}p{8.2cm}|}
    \hline

    \multicolumn{1}{|c|}{\multirow{2}{*}{\makecell{\bf Environmental\\ \bf Concern}}}&
    \multicolumn{1}{c|}{\multirow{2}{*}{\makecell{\bf Authors}}}& 
    \multicolumn{2}{c|}{\bf Implementation Type}&
    \multirow{2}{*}{\makecell{\bf Link}}\\
    \cline{3-4}

    & & \textbf{\scriptsize{Blockchain}}& \textbf{\scriptsize Smart Contract}& \\ \hline
    
    {Greenhouse gas}& {Nu{\ss}baum et al.~\cite{GW8}}& {\scriptsize{\checkmark}}& {}& {https://github.com/JCCLaude/IoT-Blockchain}\\
    \hline

    {Carbon emissions}& {Effah et al.~\cite{effah2021carbonC_CM2}}& {}& {\scriptsize{\checkmark}}& {https://github.com/De-miles1/Carbon/tree/master}\\
    \cline{2-5}

    {}& {Eckert et al.~\cite{eckert2020blockchain_CM4}}& {\scriptsize{\checkmark}}& {}& {https://github.com/LiTrans/BSMD-ML}\\
    \cline{2-5}

    {}& {Yuan et al.~\cite{yuan2018design_CM18}}& {\scriptsize{\checkmark}}& {}& {https://github.com/xisiot/HyperETS}\\
    \hline

    {Solid waste}& {Ahmad et al.~\cite{SWM18}}& {}& {\scriptsize{\checkmark}}& {https://github.com/AhmadKhalifaUniversity/Code/tree/main}\\
    \cline{2-5}

    {}& {Le et al.~\cite{SWM24}}& {\scriptsize{\checkmark}}& {}& {https://github.com/Masquerade0127/medical-blockchain}\\
    \hline

    {Plastics}&	{Alnuaimi et al.~\cite{PR21_alnuaimi2023blockchain}}& {}& {\scriptsize{\checkmark}}& {https://github.com/eimalnuaimi/RecycleChain}\\
    \hline

    {Food waste}& {Dey et al.~\cite{FW6}}& {\scriptsize{\checkmark}}& {}& {https://github.com/somdipdey/SmartNoshWaste}\\
    \cline{2-5}
    
    {}& {Baralla et al.~\cite{FW38}}& {}& {\scriptsize{\checkmark}}& {https://github.com/0xjei/SawChain}\\
    \hline

    {Water}& {Iyer et al.~\cite{WT6}}& {\scriptsize{\checkmark}}& {}& {https://github.com/sreeragiyer/Wastewater-Reuse}\\
    \cline{2-5}

    {}& {Mahmoud et al.~\cite{WT40}}& {\scriptsize{\checkmark}}& {}& {https://github.com/HaithamHmahmoud/WDSchain}\\
    \cline{2-5}

    {}& {Mughal et al.~\cite{WT58}}& {}& {\scriptsize{\checkmark}}& {https://github.com/muhammadhussainmughal/}\\
    \hline

    {Circular economy}& {Eshghie et al.~\cite{CE19}}& {\scriptsize{\checkmark}}& {}& {https://github.com/Kasche153/CircleChain}\\
    \hline

\end{tabular}%
\label{tab:github}%
\end{table*}%

As observed in the previous sections, blockchain has been adopted to digitize records, facilitate tracking and traceability of products and materials, and automate tokenization in the trading of products and services to enhance truthful monitoring of environmental variables. The immutability of records supports trust and transparency within the system. While the body of work of existing blockchain applications in waste and natural resource management is large, focused and practical work is still needed to identify the far-reaching potential of blockchain.

As consensus algorithms in blockchain implementations require intensive computation, energy efficiency is a major concern. The trade-offs between energy efficiency, performance, and security are yet to be understood.

The proliferation of sensors, IoT devices, VANETs, and video feeds is large in blockchain applications. Therefore, the amount of sensing data for the environment grows at a staggering rate. That might challenge the scalability of blockchain systems and also methods to efficiently harvest, represent, and analyze these data. Furthermore, work on data analysis might offer additional views on the impact of the use of blockchain. 

We observed that the management of carbon and plastic has attracted much attention from companies and for-profit organizations where they have come up with their blockchain-based solutions. However, their implemented systems are mostly proprietary and do not offer technical details on the blockchain they use or share data for public access. Due to the lack of information and outcome measurement, it is unclear how the environmental impact of these systems is evaluated and whether they are effective.

The use of data to trace and track materials or products owned by consumers using blockchain also may raise privacy concerns. The collection of data on the life cycle of the material may offer information on the activities of a consumer. Therefore, measures to protect the privacy of consumers while providing tracking, tracing, and transparency features, among others, to the blockchain-based management systems deserve future research.


We highlight the reported blockchain applications that not only have been practical and implemented studies but also share part source code. These implementations encompass blockchain or smart contract codes, as detailed in Table~\ref{tab:github}. While our presentation of this code does not imply endorsement (nor the opposite), we provide the associated links to facilitate thorough examination. The descriptions of the code can be found in the corresponding references.

\section{Conclusions}
\label{sec:conclusions}

We have reviewed existing blockchain applications for protecting data in the management of waste, food, and water to minimize negative impacts on environmental sustainability. We have categorized the broad area of environmental monitoring into greenhouse gas emissions, carbon, solid waste, plastic waste, water, food, and circular economy.

We identified the motivations for using blockchain in the surveyed research work and development, the groups of stakeholders that might have conflicting interests in reporting actual data, as identified in the original literature, the frameworks used and consensus algorithms, and the supporting technology. We highlighted the features that existing work aimed to achieve through the use of blockchain and unveiled the remaining challenges in each of the different application categories. 

To provide a thorough overview of the state of the art in such topics, we summarized the focus of existing surveys on these and related topics to highlight their contributions. In the end, we discuss the remaining challenges that signal the direction for future research. For practical value, we identify those works that reportedly reached implementation states and shared source code used to model blockchain artifacts used in the presented systems. Therefore, we provided answers to the proposed research questions according to recent reported work in the surveyed environmental factors and blockchain.

\section*{Acknowledgment}
This material is based upon work partially supported by the National Science Foundation under Grant No. 1856032 and NJIT 2023 Seed Grant.


\end{document}